\documentclass[english]{paper}
\usepackage[T1]{fontenc}
\usepackage[latin9]{inputenc}
\usepackage{geometry}
\usepackage{color}
\usepackage{babel}
\usepackage{float}
\usepackage{mathrsfs}
\usepackage{amsmath}
\usepackage{amssymb}
\usepackage{setspace}
\usepackage[unicode=true,pdfusetitle,
 bookmarks=true,bookmarksnumbered=false,bookmarksopen=false,
 breaklinks=true,pdfborder={0 0 0},pdfborderstyle={},backref=false,colorlinks=true]
 {hyperref}
\hypersetup{
    colorlinks=true,
    linkcolor=red,
    filecolor=magenta,      
    urlcolor=cyan,
    citecolor=blue
}
\makeatletter

\providecommand{\tabularnewline}{\\}

\numberwithin{figure}{section}
\newcommand{\lyxaddress}[1]{
	\par {\raggedright #1
	\vspace{1.4em}
	\noindent\par}
}

\@ifundefined{date}{}{\date{}}
\makeatother

\begin{document}
\title{{\Large{}Quaternionic approach on the Dirac-Maxwell, Bernoulli and
Navier-Stokes equations for dyonic fluid plasma}}
\author{\textrm{\textbf{\textup{\large{}B. C. Chanyal }}}\thanks{\textit{Email: bcchanyal@gmail.com, bcchanyal@gbpuat.ac.in}}}
\maketitle

\lyxaddress{\textit{Department of Physics, G. B. Pant University of Agriculture
\& Technology, Pantnagar-263145 (Uttarakhand), India}}
\begin{abstract}
Applying the Hamilton's quaternion algebra, we propose the generalized
electromagnetic-fluid dynamics of dyons governed by the combination
of the Dirac-Maxwell, Bernoulli and Navier-Stokes equations. The generalized
quaternionic hydro-electromagnetic field of dyonic cold plasma consist
the electrons and the magnetic monopoles in which there exist dual-mass
and dual-charge species in presence of dyons. We construct the conservation
of energy and conservation of momentum equations by equating the quaternionic
scalar and vector parts for generalized hydro-electromagnetic field
of dyonic cold plasma. We propose the quaternionic form of conservation
of energy is related to the Bernoulli's like equation while the conservation
of momentum is related to Navier-Stokes like equation for dynamics
of dyonic plasma fluid. Further, the continuity equation i.e. the
conservation of electric and magnetic charges with the dynamics of
hydro-electric and hydro-magnetic flow of conducting cold plasma fluid
is also analyzed. The quaternionic formalism for dyonic plasma wave
emphasizes that there are two types of waves propagation namely the
Langmuir like wave propagation due to electrons, and the \textquoteright t
Hooft-Polyakov like wave propagation due to magnetic monopoles.


\end{abstract}

\section{Introduction }

The plasma is the dominant constituent of matter in the universe.
The properties of the plasma is entirely different from that of the
ordinary gases and solids. Due to presence the cluster of the charged
particles in the plasma, it shows the collective behavior which represents
the long range Coulomb force between the plasma particles. There are
two types of interactions in the plasma, namely the charge-charge
interactions and charge-neutral interactions. In charge-charge interactions,
charge particles interact according to the Coulomb law, while in the
charge-neutral interactions there is the generation of the electric
polarization fields which may produce by the distortion of the neutral
atom when comes in contact with the charged particles. The range of
this polarization field is limited in the order of diameter of the
atom, i.e. it effected only for the inter-atomic distance to perturb
the orbital motion of the electrons. This interaction also involves
the induced or permanent dipole moments. Furthermore, to explore the
properties of plasma, it is important to study the influence of applied
electric and magnetic field. Due to the high mobility of the electron,
the plasmas are generally considered as the good conductor of electrical
as well as the thermal conductivity. There is the diffusion of the
charge particles in plasma from high density to the region of low
density due to the particles density gradient. The charge particle
electron, due to its lower mass and high mobility, is more diffusible
than ions. Moreover, a plasma also has the property to sustain a wave
phenomena due to the charge particles. In low-frequency region, Alfven
waves and magnetosonic waves are studied, whereas in high frequency
region longitudinal electrostatic wave and transverse electromagnetic
wave are studied. 

Many researcher studied the behavior of electrically conducting fluid
plasma in the presence of magnetic field. Alfven \cite{key-1} proposed
the theory of Magneto-hydrodynamics (MHD) and suggested that electrically
conducting fluid can support the propagation of shear waves. Meyer-Vernet\textbf{
}\cite{key-2} discussed how the electromagnetic wave propagates in
a cold plasma that contained both electric and the magnetic charges.
Correspondingly, Kambe \cite{key-3} constructed the mathematical
formulation for compressible fluids, which provides an analogous theory
of Maxwell equations for the viscous fluids. The magnetic field works
as a vorticity field whereas the electric field works as the Lamb
vector field. It shows the complete analogous theory of electromagnetism
in terms of fluid mechanics where the fluid-flow follows the Galilean
symmetry whereas electromagnetic field follows the Lorentz symmetry.
Further, Thompson-Moeller\textbf{ }\cite{key-4} also have interpreted
the Maxwell like equations for plasma particles.

In mathematical physics, the study of four dimensional particles like
dyons, tachyons, etc. in distinguish mediums can be explain by the
help of division algebras. Basically, there exist four types of divisions
algebras \cite{key-5}, they are the real, complex, quaternion and
octonion algebras. The quaternionic algebra \cite{key-6} is generally
an extension of complex numbers, can be express by the four-dimensional
Euclidean spaces \cite{key-7,key-8}. The quaternionic algebra has
vast applications in the multiple branches of theoretical physics.
The Maxwell\textquoteright s equations in the presence of magnetic
monopoles, and other classical equations of motion have already been
developed in terms of quaternionic algebra \cite{key-9}.\textbf{
}Moreover, Bisht \textit{et al}. \cite{key-10}\textbf{ }discussed
the MHD equations of plasma for massive dyons containing electric
as well as magnetic charge. Thus, keeping in mind the properties of
quaternionic algebra and its application in theoretical physics, in
this paper, we discuss the behavior of hydro-electromagnetic field
of dyonic cold plasma and its conservation laws in terms of quaternionic
field. We propose the quaternionic energy-momentum conservation laws
for dyonic plasma particle. In this case the conservation of energy
is related to the Bernoulli's like equation while the conservation
of momentum is related to Navier-Stokes like equation for dynamics
of dyonic plasma particle. Further, the quaternionic expression for
dyonic plasma wave emphasizes that there are two types of waves propagation
namely the Langmuir like wave propagation due to electrons, and the
\textquoteright t Hooft-Polyakov like wave propagation due to magnetic
monopoles. The present theory also unify the Langmuir and \textquoteright t
Hooft-Polyakov like waves in a single quaternionic framework.

\section{Preliminaries}

In microscopic description of plasma particles, we consider plasma
particles as the point-like classical particles where the quantum
effect becomes negligible. let us start with a single plasma particle
governs the spatial distribution by the Dirac delta function as \cite{key-11}
\begin{align}
\delta[\boldsymbol{r}-\boldsymbol{r}(t)]\,\,= & \,\delta[x-x(t)]\,\delta[y-y(t)]\,\delta[z-z(t)]\,,\label{eq:1}
\end{align}
where $\boldsymbol{r}\,(x,\,y,\,z)$ is fixed coordinate and $\boldsymbol{r}(t)$
is any trajectory for moving plasma particle. For this case, the velocity
space distribution in a six dimensional phase-space for plasma particle
will be $\delta[\boldsymbol{v}-\boldsymbol{v}(t)]$. The microscopic
distribution for $N$- charged particles of plasma in given phase-space
can be written as
\begin{align}
f\,(\boldsymbol{r},\,\boldsymbol{v},\,t)\,\,=\,\, & \sum_{j=1}^{N}\delta[\boldsymbol{r}-\boldsymbol{r}(t)]\,\delta[\boldsymbol{v}-\boldsymbol{v}(t)]\,,\label{eq:2}
\end{align}
where the particle density becomes
\begin{align}
\mathsf{n}\,(\boldsymbol{r},\,t)\,\,= & \,\,\int d^{3}v\,f\,(\boldsymbol{r},\,\boldsymbol{v},\,t)\,\,=\,\,\sum_{j=1}^{N}\delta[\boldsymbol{r}-\boldsymbol{r}(t)]\,.\label{eq:3}
\end{align}
The equation of motion for $j^{th}$charge particle of plasma under
the influence of Lorentz force due to the electric ($\boldsymbol{E}$)
and magnetic induction ($\boldsymbol{B}$) fields in particle trajectories
($\boldsymbol{r}_{j}(t)$, $\boldsymbol{v}_{j}(t)$) can be written
as
\begin{align}
m\frac{d\boldsymbol{v}_{j}(t)}{dt}\,\,= & \,\,q_{j}\left[\boldsymbol{E}(\boldsymbol{r}_{j},\,t)+\boldsymbol{v}_{j}\times\boldsymbol{B}((\boldsymbol{r}_{j},\,t))\right]\,\,,\label{eq:4}\\
\frac{d\boldsymbol{r}_{j}(t)}{dt}\,\,= & \,\,\boldsymbol{v}_{j}\,,\,\,\,\,(\forall\,j=1,2,.........N)\,.\label{eq:5}
\end{align}
The electric and magnetic fields satisfy the following Maxwell's equations,
\begin{align}
\boldsymbol{\nabla\cdot E}\, & =\,\rho_{c}\,,\label{eq:6}\\
\boldsymbol{\nabla\cdot B}\, & =\,0\,,\label{eq:7}\\
\boldsymbol{\nabla\times E} & \,=-\frac{\partial\boldsymbol{B}}{\partial t}\,,\label{eq:8}\\
\boldsymbol{\nabla\times B}\, & =\,\frac{\partial\boldsymbol{E}}{\partial t}+\boldsymbol{J}\,,\label{eq:9}
\end{align}
where we consider the natural unit ($\hbar=c=1$). The required charge
and current density, respectively $(\rho_{c},\,\boldsymbol{J})$ can
also be expressed as,
\begin{align}
\rho_{c}\,(\boldsymbol{r},\,t)\,\,= & \,\sum_{s}q_{s}\int d^{3}v\,f\,(\boldsymbol{r},\,\boldsymbol{v},\,t)\,\,=\,\,\sum_{s}q_{s}\sum_{j=1}^{N}\delta[\boldsymbol{r}-\boldsymbol{r}_{j}(t)]\,,\label{eq:10}\\
\boldsymbol{J}\,(\boldsymbol{r},\,t)\,\,= & \,\sum_{s}q_{s}\int d^{3}v\,\boldsymbol{v}\,f\,(\boldsymbol{r},\,\boldsymbol{v},\,t)\,\,=\,\,\sum_{s}q_{s}\sum_{j=1}^{N}\boldsymbol{v}_{j}(t)\delta[\boldsymbol{r}-\boldsymbol{r}_{j}(t)]\,,\label{eq:11}
\end{align}
where $q_{s}$ is the effective charge of $s-$species. The total
time derivation of equation (\ref{eq:2}) gives the complete microscopic
description of plasma for $s-$species \cite{key-12}
\begin{align}
\frac{df_{s}}{dt}\,\,= & \,\frac{\partial f_{s}}{\partial t}+\boldsymbol{v}\cdot\frac{\partial f_{s}}{\partial\boldsymbol{r}}+\frac{q_{s}}{m_{s}}(\boldsymbol{E}(\boldsymbol{r},\,t)+\boldsymbol{v}\times\boldsymbol{B}(\boldsymbol{r},\,t))\cdot\frac{\partial f_{s}}{\partial\boldsymbol{v}}\,=\,\,0\,.\label{eq:12}
\end{align}
This equation is called Klimontovich equations which describe the
$N-$particles motion in a single equation. The Coulomb collision
phenomena can also effect the motion of plasma particles due to its
charge dependent. But for some plasma processes we can neglect the
Coulomb collision effect. To express the collisionless plasma, the
kinetic equation can be written by using the average of Boltzmann
distribution function \cite{key-13}, i.e., $\left\langle f_{s}\right\rangle \rightarrow f$
, $\left\langle q_{s}\right\rangle \rightarrow q$, $\left\langle m_{s}\right\rangle \rightarrow m$
as,
\begin{align}
\frac{df}{dt}=\frac{\partial f}{\partial t}+\boldsymbol{v}\cdot\frac{\partial f}{\partial\boldsymbol{r}}+\frac{q}{m}(\boldsymbol{E}+\boldsymbol{v}\times\boldsymbol{B})\cdot\frac{\partial f}{\partial\boldsymbol{v}} & \,=\,\,0\,\,.\label{eq:13}
\end{align}
On the other hand, the plasma can also describe by fluid theory where
two interpenetrating fluids are electrons-fluid and ions-fluid. In
two fluid theory of plasma the continuity equations define the mass
conservation and charge conservation laws, i.e.,

\begin{align}
\frac{\partial\rho_{M}}{\partial t}+\boldsymbol{\nabla}\cdot\boldsymbol{J}_{M} & \,=\,\,0\,,\label{eq:14}\\
\frac{\partial\rho_{c}}{\partial t}+\boldsymbol{\nabla}\cdot\boldsymbol{J}_{c} & \,=\,\,0\,,\label{eq:15}
\end{align}
where the mass and charge densities $\left(\rho_{M},\,\rho_{c}\right)$
are given as
\begin{align}
\rho_{M}\,\,= & \,\,\,m_{e}n_{e}+m_{i}n_{i}\,,\label{eq:16}\\
\rho_{c}\,\,= & \,\,\,q_{e}n_{e}+q_{i}n_{i}\,.\label{eq:17}
\end{align}
Here $\left(m_{e},\,n_{e},\,q_{e}\right)$ and $\left(m_{i},\,n_{i},\,q_{i}\right)$
are the electronic fluid and ionic fluid terms respectively for masses,
total number of particles and charges. Similarly, the current sources
$\left(\boldsymbol{J}_{M},\,\boldsymbol{J}_{c}\right)$ due to masses
and charges of two-fluid plasma can be written as
\begin{align}
\boldsymbol{J}_{M}\,\,= & \,\,\rho_{M}\boldsymbol{v}\,\,=\,\,m_{e}n_{e}\boldsymbol{v}_{e}+m_{i}n_{i}\boldsymbol{v}_{i}\,,\label{eq:18}\\
\boldsymbol{J}_{c}\,\,= & \,\,\rho_{c}\boldsymbol{v}\,\,=\,\,\,q_{e}n_{e}\boldsymbol{v}_{e}+q_{i}n_{i}\boldsymbol{v}_{i}\,\,,\label{eq:19}
\end{align}
where the center of mass fluid velocity $\boldsymbol{v}$ will be
\begin{align}
\boldsymbol{v}\,\,= & \,\,\,\frac{1}{\rho_{M}}\left(\boldsymbol{v}_{e}m_{e}n_{e}+\boldsymbol{v}_{i}m_{i}n_{i}\right)\,.\label{eq:20}
\end{align}
Another equation to the fluid theory is force equation that gives
the exact motion of plasma fluid. This can be written as 
\begin{align}
m_{s}\frac{d\boldsymbol{v}_{s}(\boldsymbol{r},\,t)}{dt}\,\,= & \,\,F_{s}(\boldsymbol{r},\,t)\,,\label{eq:21}
\end{align}
where $F_{s}(\boldsymbol{r},\,t)$ is the total force per unit volume
acting on the fluid species at space-time $(\boldsymbol{r},\,t)$
and the acceleration of conducting fluid species yield
\begin{align}
\frac{d\boldsymbol{v}_{s}(\boldsymbol{r},\,t)}{dt}\,\,=\,\, & \left(\frac{\partial}{\partial t}+\boldsymbol{v}_{s}\cdot\boldsymbol{\nabla}\right)\boldsymbol{v}_{s}\,.\label{eq:22}
\end{align}
Here, the term $\left(\boldsymbol{v}_{s}\cdot\boldsymbol{\nabla}\right)\boldsymbol{v}_{s}$
used for the convective acceleration of fluid particles. In given
equation (\ref{eq:21}) the total force acting on the plasma fluid
species may be the resultant of the pressure gradient force and the
Lorentz electromagnetic force. Therefore,
\begin{align}
\rho_{M}\left(\frac{\partial}{\partial t}+\boldsymbol{v}_{s}\cdot\boldsymbol{\nabla}\right)\boldsymbol{v}_{s}\,\,= & -\boldsymbol{\nabla}p_{s}+\frac{q_{s}}{m_{s}}\boldsymbol{E}+\frac{q_{s}}{m_{s}}(\boldsymbol{v}_{s}\times\boldsymbol{B})\,,\label{eq:23}
\end{align}
where $\boldsymbol{\nabla}p_{s}$ indicated the pressure force acting
due to the inhomogeneity of the plasma. The generalized Ohm's law
\cite{key-13} for plasma fluid species can also be written as
\begin{align}
\frac{m_{e}m_{i}}{\rho_{M}\,e^{2}}\frac{\partial\boldsymbol{J}_{c}}{\partial t} & \,\,=\,\,\frac{m_{i}}{2\rho_{M}\,e}\boldsymbol{\nabla}p_{e,i}+\boldsymbol{E}+\left(\boldsymbol{v}_{e,i}\times\boldsymbol{B}\right)-\frac{m_{i}}{\rho_{M}\,e}\left(\boldsymbol{J}_{c}\times\boldsymbol{B}\right)-\frac{\boldsymbol{J}_{c}}{\sigma}\,,\label{eq:24}
\end{align}
where $\sigma$ introduced for the conductivity of plasma fluid. If
we combine the conducting plasma fluid with electromagnetic field
then the relevant fluid theory called MHD \cite{key-14}. In MHD,
the simplest system for macroscopic transport equations of fluid plasma
is known as the cold plasma model. We introduce the following approximation
of fluid parameters to the case of cold plasma \cite{key-15,key-16,key-17}
\begin{align}
T_{e,i}\,\,\, & \sim\,\,\,0,\,\,\,\,\,\,\,\,\,\,\,\,\,\,\boldsymbol{\nabla}p\,\,\sim\,\,\,0\,,\nonumber \\
\mathscr{E}_{e}\,\,\, & \sim\,\,\,\mathscr{E}_{i}\,,\,\,\,\,\,\,\,\,\,\,\,\,\boldsymbol{v}_{e}\,\,\,\sim\,\,\,\boldsymbol{v}_{i}\,,\label{eq:25}\\
\rho_{e}\,\,\, & \sim\,\,\,\rho_{i}\,,\,\,\,\,\,\,\,\,\,\,\,\,\,\,n_{e}\,\,\,\sim\,\,\,n_{i}\,.\nonumber 
\end{align}
Here $T$ is temperature and $\mathscr{E}$ is effective energy of
fluid particles. Therefore, the Navier-Stokes and continuity equation
for cold plasma fluid yield
\begin{align}
\rho\left(\frac{\partial}{\partial t}+\boldsymbol{v}\cdot\boldsymbol{\nabla}\right)\boldsymbol{v}\,\,= & \,\,\frac{q}{m}\left[\boldsymbol{E}+(\boldsymbol{v}\times\boldsymbol{B})\right]\,,\label{eq:26}
\end{align}

\begin{align}
\frac{\partial\rho}{\partial t}+\boldsymbol{\nabla}\cdot\boldsymbol{J} & \,=\,\,0\,,\label{eq:27}
\end{align}
where $\rho$ is cold mass density, $\boldsymbol{v}$ is cold fluid
velocity, $m$ is fluid mass, and $q$ is cold charge. As such the
Ohm's law associated with cold current source $\boldsymbol{J}$ as
\begin{align}
\frac{m^{2}}{\rho\,e^{2}}\frac{\partial\boldsymbol{J}}{\partial t} & \,\,=\,\,\boldsymbol{E}+\left(\boldsymbol{v}\times\boldsymbol{B}\right)-\frac{m}{\rho\,e}\left(\boldsymbol{J}\times\boldsymbol{B}\right)-\frac{\boldsymbol{J}}{\sigma}\,.\label{eq:28}
\end{align}
Equations (\ref{eq:26})-(\ref{eq:28}) in the cold plasma shows the
temperature independent dispersion relation or we can say that, the
thermal velocity of the particle is small to compared with the wave
phase velocity. Basically, in cold plasma approximation we are not
considered the individual motion of the electrons or ions. Here, we
take equivalent motion of electrons and ions to case of cold plasma-fluid
approximation with temperature $T=0$.

\section{The quaternionic field}

The hyper-complex algebras are widely used to explain many theories
\cite{key-18}-\cite{key-26} related to high energy physics. In hyper-complex
algebras, quaternion is a four dimensional norm-division algebra over
the field of real numbers $\mathbb{R}$ invented by Hamilton \cite{key-6}.
A quaternionic variable ($\mathbb{Q}$) can be expressed by the unification
of scalar as well as vector spaces, i.e.,
\begin{align}
\mathbb{Q}\,\,= & \,\,\left(q_{0},\,\boldsymbol{q}\right)\,\,\,\simeq\,\,\,S(q)+\boldsymbol{V}(q)\,,\,\,\,\,\forall\,\mathbb{Q}\in\mathbb{H}\,,\nonumber \\
= & \,\,e_{0}q_{0}+\sum_{j=1}^{3}e_{j}q_{j}\,,\,\,\,\left(\forall\,q_{0}\in\mathbb{R},\,\,q_{j}\in\mathbb{R}^{3}\right)\,\,,\label{eq:29}
\end{align}
where $(S(q))$ is the scalar and $(\boldsymbol{V}(q))$ is the vector
field in Hamilton space ($\mathbb{H}$) associated with quaternionic
unit elements ($e_{0}$, $e_{1}$, $e_{2}$, $e_{3}$). The quaternionic
conjugate $\bar{\mathbb{Q}}$ in the same $\mathbb{H}$-space can
be written as
\begin{align}
\bar{\mathbb{Q}}\,\,= & \,\,S(q)-\boldsymbol{V}(q)\,\,=\,\,\,\,e_{0}q_{0}-\sum_{j=1}^{3}e_{j}q_{j}\,\,.\label{eq:30}
\end{align}
From equations (\ref{eq:29}) and (\ref{eq:30}) we also can define
the real and imaginary quaternions, viz. $\text{Re}\,(\mathbb{H}):\longmapsto q_{0}$$=(q+\bar{q})/2$
and $\text{Im}\,(\mathbb{H}):\longmapsto q_{j}$ $=(q-\bar{q})/2$.
The quaternionic basis vectors satisfy the given multiplication rules
\begin{align}
e_{0}e_{0}\, & =\,\,e_{0}^{2}=\,1\,,\,\,e_{A}^{2}=\,-1\,,\,\,e_{0}e_{A}=\,e_{A}e_{0}=e_{A}\,,\nonumber \\
e_{A}e_{B} & =\,\,-\delta_{AB}e_{0}+f_{ABC}e_{C}\,,\,\,\,\,\,\,\,(\forall\,A,B,C=1,2,3)\,,\label{eq:31}
\end{align}
where $\delta_{AB}$ and $f_{ABC}$ are delta symbol and Levi Civita
symbol, respectively. As such, the commutation and anti-commutation
relations for quaternionic basis vectors are expressed as
\begin{align}
\left[e_{A},\,\,e_{B}\right] & \,=\,2\,f_{ABC}\,e_{C}\,,\,\,\,\,\,\,\,\,\,\,\,\,(\text{commutation relation})\label{eq:32}\\
\left\{ e_{A},\,\,e_{B}\right\}  & \,=\,-2\,\delta_{AB}e_{0}\,,\,\,\,\,\,\,\,\,(\text{anti-commutation relation})\,.\label{eq:33}
\end{align}
The quaternion holds the associative law, i.e.,
\begin{align}
e_{A}(\,e_{B}\,e_{C}) & \,=\,(e_{A}\,e_{B}\,)\,e_{C}\,.\label{eq:34}
\end{align}
The addition and the multiplication of any two quaternions are expressed
by
\begin{align}
\mathbb{Q}\pm\mathbb{P}\,\,= & \,\left(q_{0}\pm p_{0}\right)+\left(\boldsymbol{q}\pm\boldsymbol{p}\right)\nonumber \\
= & \,e_{0}\left(q_{0}\pm p_{0}\right)+e_{1}\left(q_{1}\pm p_{1}\right)+e_{2}\left(q_{1}\pm p_{1}\right)+e_{3}\left(q_{1}\pm p_{1}\right)\,\,,\label{eq:35}
\end{align}
\begin{align}
\mathbb{Q}\circ\mathbb{P}\,\,= & \,\left[q_{0}+\boldsymbol{q}\right]\left[p_{0}+\boldsymbol{p}\right]\nonumber \\
= & \,\,e_{0}(q_{0}p_{0}-\boldsymbol{q}\cdot\boldsymbol{p})+e_{j}\left(q_{0}\boldsymbol{p}+p_{0}\boldsymbol{q}+(\boldsymbol{q}\times\boldsymbol{p})\right)\,\,,\,\,(\forall\,j=1,2,3)\,\,,\label{eq:36}
\end{align}
where we notice that the quaternionic multiplication is non-commutative,
i.e., $\mathbb{Q}\circ\mathbb{P}\,\neq\,\mathbb{P}\circ\mathbb{Q}$,
because $\boldsymbol{q}\times\boldsymbol{p}\neq0$ and $\boldsymbol{q}\times\boldsymbol{p}\neq\boldsymbol{p}\times\boldsymbol{q}$.
Further, the quaternionic Euclidean scalar product $\mathbb{H}\times\mathbb{H}\longmapsto\mathbb{R}$
can also be written as
\begin{align}
\left\langle \mathbb{Q},\,\mathbb{P}\right\rangle \,=\,\,\text{Re}\,(\mathbb{Q}\circ\bar{\mathbb{P}}) & \,=\,\left(q_{0}p_{0}+q_{1}p_{1}+q_{2}p_{2}+q_{3}p_{3}\right)\,.\label{eq:37}
\end{align}
The quaternionic modulus $\mid\mathbb{Q}\mid$ and quaternionic inverse
$\mathbb{Q}^{-1}$ are respectively expressed by
\begin{align}
\mid\mathbb{Q}\mid\,= & \,\sqrt{q_{0}^{2}+q_{1}^{2}+q_{2}^{2}+q_{3}^{2}}\,,\label{eq:38}\\
\mathbb{Q}^{-1}\,= & \,\frac{\bar{q}}{\mid q\mid}\,.\label{eq:39}
\end{align}
The multiplication rules for quaternion conjugation and norm are given
as,
\begin{align}
\overline{\mathbb{Q}_{1}\circ\mathbb{Q}_{2}}\,\,= & \,\,\overline{\mathbb{Q}_{1}}\,\circ\,\overline{\mathbb{Q}_{2}}\,\,\label{eq:40}
\end{align}
\begin{align}
N\left(\mathbb{Q}_{1}\circ\mathbb{Q}_{2}\right)\,= & \,N\left(\mathbb{Q}_{1}\right)\,\circ\,N\left(\mathbb{Q}_{2}\right)\,\,.\label{eq:41}
\end{align}
The quaternion unit elements show non-Abelian structure in nature
and thus follow the non-commutative division ring. Moreover, in the
application of physics Girard \cite{key-27} discussed the role of
quaternionic group in modern physics, i.e., the effect of quaternions
in SO(3), the Clifford algebra SU(2), the Lorentz group and the conformal
group. Recently, quaternionic formulation has been applied to describe
the quantized equation of electromagnetism of dyons \cite{key-28,key-29}. 

\section{Generalized dual MHD of cold plasma in Hamilton space}

The dual MHD field consist not only electrons and ions but also consist
with the magnetic monopoles and their ionic partners magneto-ions
\cite{key-30}. To study the dyonic cold plasma field, there are dual-mass
and dual-charge species in presence of dyons. Many authors \cite{key-31,key-32,key-33}
discussed the generalized fields associated with dyons. From equation
(\ref{eq:25}), we consider electrons and magnetic monopoles (constitute
of \textit{dyons}) are equivalent to ions and magneto-ions (constitute
of \textit{i-dyons}) in cold plasma approximation. Therefore, the
dyonic equivalent of cold plasma equations are written as the following
ways:
\begin{align}
\varrho^{D}(\varrho^{e},\,\varrho^{\mathfrak{m}})\,\simeq\, & \left(m^{e}n^{e}+m^{\mathfrak{m}}n^{\mathfrak{m}}\right)\,,\,\,\,\,\,\,\,(\text{dual-mass density})\label{eq:42}\\
\rho^{D}(\rho^{e},\,\rho^{\mathfrak{m}})\,\simeq\, & \left(q^{e}n^{e}+q^{\mathfrak{m}}n^{\mathfrak{m}}\right)\,,\,\,\,\,\,\,\,(\text{dual-charge density})\label{eq:43}\\
\boldsymbol{v}^{D}(\boldsymbol{v}^{e},\,\boldsymbol{v}^{\mathfrak{m}})\,\,\simeq & \,\,\,\frac{1}{\varrho^{D}}\left(\boldsymbol{v}^{e}m^{e}n^{e}(x)+\boldsymbol{v}^{\mathfrak{m}}m^{\mathfrak{m}}n^{\mathfrak{m}}(x)\right)\,,\,\,\,\,\,\,\,(\text{dual-mass velocity})\label{eq:44}\\
\frac{\partial\varrho^{D}}{\partial t}+\boldsymbol{\nabla}\cdot(\varrho^{D}\boldsymbol{v}^{D})\, & =\,\,0\,,\,\,\,\,\,\,\,(\text{dual-mass conservation law})\label{eq:45}\\
\frac{\partial\rho^{D}}{\partial t}+\boldsymbol{\nabla}\cdot\boldsymbol{J}^{D}\, & =\,\,0\,,\,\,\,\,\,\,\,(\text{dual-charge conservation law})\label{eq:46}
\end{align}
where $(\varrho^{e},\,\varrho^{\mathfrak{m}})$ and $(\rho^{e}$,$\rho^{\mathfrak{m}})$
are the electric, magnetic mass and charge densities respectively,
while $(\boldsymbol{J}^{e}=\,q^{e}n^{e}\boldsymbol{v}^{e}$,~$\boldsymbol{J}^{\mathfrak{m}}=\,q^{\mathfrak{m}}n^{\mathfrak{m}}\boldsymbol{v}^{\mathfrak{m}})$
are the two current densities associated with electric and magnetic
charges of dyons. Similarly, $(m^{e},\,n^{e},\,q^{e})$ and $(m^{\mathfrak{m}},\,n^{\mathfrak{m}},\,q^{\mathfrak{m}})$
are the mass, total number and charge for electrons and magnetic monopoles,
respectively. The dual Lorentz force equation for dyons can also be
expressed as
\begin{align}
\boldsymbol{F}^{D}\,\,= & \,\,\rho^{e}\boldsymbol{E}+\left(\boldsymbol{J}^{e}\boldsymbol{\times B}\right)+\rho^{\mathfrak{m}}\boldsymbol{B}-\left(\boldsymbol{J}^{\mathfrak{m}}\boldsymbol{\times E}\right)\,,\label{eq:47}
\end{align}
where we neglect the dyonic pressure gradient term $\left(\boldsymbol{\nabla}p\right)^{D}$
to considering cold plasma approximation \cite{key-17}. The above
equations (\ref{eq:42})-(\ref{eq:47}) are well known equations for
dual field of massive dyons. In order to discuss the quaternionic
space-time revolution of these dual field equations for cold dyonic
fluid plasma, let us write the quaternionic-valued differential operator
and its quaternionic conjugate as

\begin{align}
\mathrm{\mathbb{D\,}}(e_{1},e_{2},e_{3};\,\,\,e_{0})\,\,= & \,\,\left(e_{\mathrm{1}}\frac{\partial}{\partial x}+e_{2}\frac{\partial}{\partial y}+e_{3}\frac{\partial}{\partial z}\right)-e_{0}\frac{i}{a_{0}}\frac{\partial}{\partial t}\,,\label{eq:48}
\end{align}

\begin{align}
\bar{\mathbb{D}}\,(e_{1},e_{2},e_{3};\,\,\,e_{0})\,\,= & \,\,-\left(e_{\mathrm{1}}\frac{\partial}{\partial x}+e_{2}\frac{\partial}{\partial y}+e_{3}\frac{\partial}{\partial z}\right)-e_{0}\frac{i}{a_{0}}\frac{\partial}{\partial t}\,,\label{eq:49}
\end{align}
where $a_{0}$ denoted the speed of fluid particles. The D' Alembert
operator $\square$, can be expressed as
\begin{alignat}{1}
\square\,\,\longmapsto\,\,\,\mathbb{\left(D\circ\bar{D}\right)}\,\,\,= & \,\,\,\frac{\partial^{2}}{\partial x^{2}}+\frac{\partial^{2}}{\partial y^{2}}+\frac{\partial^{2}}{\partial z^{2}}-\frac{1}{a_{0}^{2}}\frac{\partial^{2}}{\partial t^{2}}\nonumber \\
= & \,\,\,\boldsymbol{\nabla}^{2}-\frac{1}{a_{0}^{2}}\frac{\partial^{2}}{\partial t^{2}}\,\,\simeq\,\,\mathbb{\bar{D}\circ D}\,.\label{eq:50}
\end{alignat}
In generalized MHD-field of dyonic particles, the quaternionic valued
dual-velocities can be written as
\begin{align}
\mathbb{U}\left(e_{1},\,e_{2},\,e_{3};\,\,\,e_{0}\right)\,= & \,\left\{ u_{x},\,u_{y},\,u_{z};\,\,-\frac{i}{a_{0}}h\right\} \,,\label{eq:51}\\
\mathbb{V}\left(e_{1},\,e_{2},\,e_{3};\,\,\,e_{0}\right)\,= & \,\left\{ \upsilon_{x},\,\upsilon_{y},\,\upsilon_{z};\,\,-ia_{0}k\right\} \,,\label{eq:52}
\end{align}
where $\mathbb{U}$ represents the four-components quaternionic velocity
of electrons while $\mathbb{V}$ represents the four-components quaternionic
velocity of magnetic-monopoles. Due to their mass variation the velocities
are taken different. The scalar components ($h$, $k$) represent
the two-enthalpy of dyons. Now, the bi-quaternionic (complex quaternion)
generalization of dyonic fluid velocity $\mathbb{W}$ can be written
as
\begin{align}
\mathbb{W}\,\left(e_{1},\,e_{2},\,e_{3};\,\,\,e_{0}\right)= & \,\,\left(\mathbb{U}-\frac{i}{a_{0}}\mathbb{V}\right)\nonumber \\
= & \,\,\,e_{1}\left(u_{x}-\frac{i}{a_{0}}\upsilon_{x}\right)+e_{2}\left(u_{y}-\frac{i}{a_{0}}\upsilon_{y}\right)+e_{3}\left(u_{z}-\frac{i}{a_{0}}\upsilon_{z}\right)-\frac{i}{a_{0}}e_{0}(h-ia_{0}k)\,.\label{eq:53}
\end{align}
Using equations (\ref{eq:48}) and (\ref{eq:53}), we can write the
quaternionic hydrodynamics field equation for dyonic fluid plasma
\begin{align}
\mathbb{D\,\circ W}\,\,= & \,\,\boldsymbol{\Psi}\,\,\left(e_{1},\,e_{2},\,e_{3};\,\,\,e_{0}\right)\nonumber \\
\simeq & \,\,\,e_{1}\left(B_{x}+\frac{i}{a_{0}}E_{x}\right)+e_{2}\left(B_{y}+\frac{i}{a_{0}}E_{y}\right)+e_{3}\left(B_{z}+\frac{i}{a_{0}}E_{z}\right)-e_{0}\left(B_{0}-\frac{i}{a_{0}}E_{0}\right)\,,\label{eq:54}
\end{align}
where $\boldsymbol{\Psi}$ is quaternionic generalized hydro-electromagnetic
(HEM) field for dyonic cold plasma. One can defined the components
of bi-quaternionic hydrodynamics field as
\begin{align}
\psi_{1}^{\text{HEM}}:\longmapsto\left[B_{x}+\frac{i}{a_{0}}E_{x}\right]\,= & \,\left\{ \left(\boldsymbol{\nabla}\times\boldsymbol{u}\right)_{x}-\frac{1}{a_{0}^{2}}\frac{\partial\upsilon_{x}}{\partial t}-\frac{\partial k}{\partial x}\right\} +\frac{i}{a_{0}}\left\{ -\left(\boldsymbol{\nabla}\times\boldsymbol{\upsilon}\right)_{x}-\frac{\partial u_{x}}{\partial t}-\frac{\partial h}{\partial x}\right\} \,,\label{eq:55}\\
\psi_{2}^{\text{HEM}}:\longmapsto\left[B_{y}+\frac{i}{a_{0}}E_{y}\right]\,= & \,\left\{ \left(\boldsymbol{\nabla}\times\boldsymbol{u}\right)_{y}-\frac{1}{a_{0}^{2}}\frac{\partial\upsilon_{y}}{\partial t}-\frac{\partial k}{\partial y}\right\} +\frac{i}{a_{0}}\left\{ -\left(\boldsymbol{\nabla}\times\boldsymbol{\upsilon}\right)_{y}-\frac{\partial u_{y}}{\partial t}-\frac{\partial h}{\partial y}\right\} \,,\label{eq:56}\\
\psi_{3}^{\text{HEM}}:\longmapsto\left[B_{z}+\frac{i}{a_{0}}E_{z}\right]\,= & \,\left\{ \left(\boldsymbol{\nabla}\times\boldsymbol{u}\right)_{z}-\frac{1}{a_{0}^{2}}\frac{\partial\upsilon_{z}}{\partial t}-\frac{\partial k}{\partial z}\right\} +\frac{i}{a_{0}}\left\{ -\left(\boldsymbol{\nabla}\times\boldsymbol{\upsilon}\right)_{z}-\frac{\partial u_{z}}{\partial t}-\frac{\partial h}{\partial z}\right\} \,,\label{eq:57}\\
\psi_{0}^{\text{HEM}}:\longmapsto\left[B_{0}-\frac{i}{a_{0}}E_{0}\right]\,= & \left\{ \left(\boldsymbol{\nabla}\cdotp\boldsymbol{u}+\frac{1}{a_{0}^{2}}\frac{\partial h}{\partial t}\right)-\frac{i}{a_{0}}\left(\boldsymbol{\nabla}\cdotp\boldsymbol{\upsilon}+\frac{\partial k}{\partial t}\right)\right\} \,.\label{eq:58}
\end{align}
The hydro-electric field vector ($\boldsymbol{E}$) is identical to
generalized Lamb vector field while the hydro-magnetic field vector
($\boldsymbol{B}$) is identical to generalized vorticity field \cite{key-34,key-35,key-36}
for dyonic fluid plasma. The Lorenz gauge conditions may equivalent
to the continuity like equations in dyonic fluid plasma, i.e., the
scalar component $\psi_{0}^{\text{HEM}}\simeq\left[B_{0}-\frac{i}{a_{0}}E_{0}\right]=0$
become
\begin{alignat}{1}
\boldsymbol{\nabla}\cdotp\boldsymbol{u}+\frac{1}{a_{0}^{2}}\frac{\partial h}{\partial t}\, & =\,\,0\,,\label{eq:59}\\
\boldsymbol{\nabla}\cdotp\boldsymbol{\upsilon}+\frac{\partial k}{\partial t}\, & =\,\,0\,.\label{eq:60}
\end{alignat}
Equations (\ref{eq:59}) and (\ref{eq:60}) represent the condition
for the dynamics of compressible fluid where the divergence of two-fluid
velocities are not equal to zero. Thus, these equations lead to the
non-conservation form of two-enthalpy. We can summarize the quaternionic
hydro-electromagnetic field equations (i.e., dual field $(\psi_{j},\,\chi_{j})$
for $j=0,1,2,3$) in Table-1. \\
\begin{table}[H]
\begin{doublespace}
\begin{centering}
\begin{tabular}{ccc}
\hline 
\textbf{Lamb field components} & \textbf{Vorticity field components} & \textbf{Corresponding $\mathbb{Q}$-field}\tabularnewline
\hline 
\hline 
$\psi_{1}:\longmapsto\left(\boldsymbol{\nabla}\times\boldsymbol{u}\right)_{x}-\frac{1}{a_{0}^{2}}\frac{\partial\upsilon_{x}}{\partial t}-\frac{\partial k}{\partial x}$ & $\chi_{1}:\longmapsto-\left(\boldsymbol{\nabla}\times\boldsymbol{\upsilon}\right)_{x}-\frac{\partial u_{x}}{\partial t}-\frac{\partial h}{\partial x}$ & $e_{1}(\psi_{1}+\frac{i}{a_{0}}\chi_{1})$\tabularnewline
$\psi_{2}:\longmapsto\left(\boldsymbol{\nabla}\times\boldsymbol{u}\right)_{y}-\frac{1}{a_{0}^{2}}\frac{\partial\upsilon_{y}}{\partial t}-\frac{\partial k}{\partial y}$ & $\chi_{2}:\longmapsto-\left(\boldsymbol{\nabla}\times\boldsymbol{\upsilon}\right)_{y}-\frac{\partial u_{y}}{\partial t}-\frac{\partial h}{\partial y}$ & $e_{2}(\psi_{2}+\frac{i}{a_{0}}\chi_{2})$\tabularnewline
$\psi_{3}:\longmapsto\left(\boldsymbol{\nabla}\times\boldsymbol{u}\right)_{z}-\frac{1}{a_{0}^{2}}\frac{\partial\upsilon_{z}}{\partial t}-\frac{\partial k}{\partial z}$ & $\chi_{3}:\longmapsto-\left(\boldsymbol{\nabla}\times\boldsymbol{\upsilon}\right)_{z}-\frac{\partial u_{z}}{\partial t}-\frac{\partial h}{\partial z}$ & $e_{3}(\psi_{3}+\frac{i}{a_{0}}\chi_{3})$\tabularnewline
$\psi_{0}:\longmapsto\boldsymbol{\nabla}\cdotp\boldsymbol{u}+\frac{1}{a_{0}^{2}}\frac{\partial h}{\partial t}\,=\,\,0\,,$ & $\chi_{0}:\longmapsto\boldsymbol{\nabla}\cdotp\boldsymbol{\upsilon}+\frac{\partial k}{\partial t}\,=\,\,0$ & $e_{0}(\psi_{0}-\frac{i}{a_{0}}\chi_{0})$\tabularnewline
\hline 
\end{tabular}
\par\end{centering}
\end{doublespace}
\centering{}\caption{Quaternionic generalization of Lamb-vorticity field components in
presence of dyons }
\end{table}

In order to find the field source equations for dyonic plasma fluid,
we may operate $\bar{\mathbb{D}}$ on hydro-electromagnetic field
$\boldsymbol{\Psi}$ and obtain
\begin{align}
\bar{\mathbb{D}}\circ\boldsymbol{\Psi}\,\,= & \,-\mathbb{J}\,\left(e_{1},\,e_{2},\,e_{3};\,\,\,e_{0}\right)\,,\nonumber \\
= & \,\,\mu\left(e_{1}J_{x}^{e}+e_{2}J_{y}^{e}+e_{3}J_{z}^{e}-e_{0}\rho^{\mathfrak{m}}\right)-\frac{i}{a_{0}\epsilon}\left(e_{1}J_{x}^{\mathfrak{m}}+e_{2}J_{y}^{\mathfrak{m}}+e_{3}J_{z}^{\mathfrak{m}}+\rho^{e}\right)\,,\label{eq:61}
\end{align}
so that the bi-quaternionic components of dyonic plasma source are
expressed as

\begin{alignat}{1}
\mathcal{J}_{1}^{\text{HEM}}:\longmapsto\left[\mu J_{x}^{e}-\frac{i}{a_{0}\epsilon}J_{x}^{\mathfrak{m}}\right]\,\,= & \,\,\,\left[\left\{ \left(\boldsymbol{\nabla}\times\boldsymbol{B}\right)_{x}-\frac{1}{a_{0}^{2}}\frac{\partial E_{x}}{\partial t}\right\} +\frac{i}{a_{0}}\left\{ \left(\boldsymbol{\nabla}\times\boldsymbol{E}\right)_{x}+\frac{\partial B_{x}}{\partial t}\right\} \right]\,,\label{eq:62}\\
\mathcal{J}_{2}^{\text{HEM}}:\longmapsto\left[\mu J_{y}^{e}-\frac{i}{a_{0}\epsilon}J_{y}^{\mathfrak{m}}\right]\,\,= & \,\,\,\left[\left\{ \left(\boldsymbol{\nabla}\times\boldsymbol{B}\right)_{y}-\frac{1}{a_{0}^{2}}\frac{\partial E_{y}}{\partial t}\right\} +\frac{i}{a_{0}}\left\{ \left(\boldsymbol{\nabla}\times\boldsymbol{E}\right)_{y}+\frac{\partial B_{y}}{\partial t}\right\} \right]\,,\label{eq:63}\\
\mathcal{J}_{3}^{\text{HEM}}:\longmapsto\left[\mu J_{z}^{e}-\frac{i}{a_{0}\epsilon}J_{z}^{\mathfrak{m}}\right]\,\,= & \,\,\,\left[\left\{ \left(\boldsymbol{\nabla}\times\boldsymbol{B}\right)_{z}-\frac{1}{a_{0}^{2}}\frac{\partial E_{z}}{\partial t}\right\} +\frac{i}{a_{0}}\left\{ \left(\boldsymbol{\nabla}\times\boldsymbol{E}\right)_{z}+\frac{\partial B_{z}}{\partial t}\right\} \right]\,,\label{eq:64}\\
\mathcal{J}_{0}^{\text{HEM}}:\longmapsto\left[\mu\rho^{\mathfrak{m}}-\frac{i}{a_{0}\epsilon}\rho^{e}\right]\,\,= & \,\,\,\left[\boldsymbol{\nabla}\cdotp\boldsymbol{B}-\frac{i}{a_{0}}\boldsymbol{\nabla}\cdotp\boldsymbol{E}\right]\,,\label{eq:65}
\end{alignat}
where ($\boldsymbol{J}^{e}$, $\rho^{e}$) represent the electric
source current and source density while ($\boldsymbol{J}^{\mathfrak{m}}$,$\rho^{\mathfrak{m}}$)
represent the magnetic source current and source density for dyonic
fluid plasma, and ($\epsilon,\,\mu$) define the permittivity and
permeability. Interestingly, the complex-quaternionic form of dyonic
source equations reduce to
\begin{align}
\mathcal{\boldsymbol{J}}^{\text{HEM}}= & \,\,\,\left[\left\{ \left(\boldsymbol{\nabla}\times\boldsymbol{B}\right)-\frac{1}{a_{0}^{2}}\frac{\partial\boldsymbol{E}}{\partial t}\right\} +\frac{i}{a_{0}}\left\{ \left(\boldsymbol{\nabla}\times\boldsymbol{E}\right)+\frac{\partial\boldsymbol{B}}{\partial t}\right\} \right]\,,\,\,\,\,\,\,(\text{dyonic source current})\label{eq:66}\\
\rho^{\text{HEM}}= & \,\,\,\left[\boldsymbol{\nabla}\cdotp\boldsymbol{B}-\frac{i}{a_{0}}\boldsymbol{\nabla}\cdotp\boldsymbol{E}\right]\,,\,\,\,\,\,\,(\text{dyonic source density)}\,\,.\label{eq:67}
\end{align}
 Then, equation (\ref{eq:61}) leads to the following relations
\begin{align}
\boldsymbol{\nabla}\cdotp\boldsymbol{E}\,\,= & \,\,\frac{\rho^{e}}{\epsilon}\,\,,\,\,\,\,\boldsymbol{\nabla}\cdotp\boldsymbol{B}\,\,=\,\,\mu\rho^{\mathfrak{m}}\,\,,\label{eq:68}\\
\boldsymbol{\nabla}\times\boldsymbol{E}\,=\,-\frac{\partial\boldsymbol{B}}{\partial t} & -\frac{1}{\epsilon}\boldsymbol{J}^{\mathfrak{m}}\,\,,\,\,\,\,\boldsymbol{\nabla}\times\boldsymbol{B}\,=\,\frac{1}{a_{0}^{2}}\frac{\partial\boldsymbol{E}}{\partial t}+\mu\boldsymbol{J}^{e}\,\,.\label{eq:69}
\end{align}
Equations (\ref{eq:68}) and (\ref{eq:69}) represent the generalized
Dirac-Maxwell equations for hydro-electromagnetic fields of dyonic
cold plasma. These equations incorporate all dyonic particles motion
in cold plasma fluid. But these equations are incomplete to describe
dyonic cold plasma. When combine these generalized Dirac-Maxwell equations
with the Bernoulli, Navier-Stokes and continuity equations, then the
plasma fluid equations provide a complete description of quaternionic
MHD. Therefore, in next sections, we shall discuss the quaternionic
form of Bernoulli, Navier-Stokes and continuity equations for cold
dyonic plasma fluid.

\section{Generalized quaternionic Bernoulli and Navier-Stokes like equation}

The Bernoulli and Navier-Stokes equations are basically the fundamental
differential equations that describe the conservation of energy and
the conservation of momentum to the flow of fluid \cite{key-37}.
In order to derive the quaternionic Bernoulli and Navier-Stokes like
force equation for dyonic cold plasma fluid, we may operate left by
$\boldsymbol{\bar{\Psi}}$ in the quaternionic field equation (\ref{eq:61})
as
\begin{alignat}{1}
\boldsymbol{\bar{\Psi}}\circ(\mathbb{\bar{D}}\circ\boldsymbol{\Psi})\,\,= & \,-\boldsymbol{\bar{\Psi}}\circ\mathbb{J}\,.\label{eq:70}
\end{alignat}
Now, we simplify the left hand part of quaternionic field equation
(\ref{eq:70}) as
\begin{alignat}{1}
\boldsymbol{\bar{\Psi}}\circ(\mathbb{\bar{D}}\circ\boldsymbol{\Psi})\,\,\left\{ e_{1},\,e_{2},\,e_{3};\,\,\,e_{0}\right\} \,\,= & \,\,\ensuremath{e_{1}}L+\ensuremath{e_{2}}M+\ensuremath{e_{3}}N+\ensuremath{e_{0}}O\,\,,\,\,\,\forall\,(L,\,M,\,N,\,O)\in\mathbb{C},\label{eq:71}
\end{alignat}
where the real and imaginary quaternionic components $(L,M,N,\,\text{and\,}O)$
are expressed by
\begin{alignat}{1}
\text{Re}\left\{ e_{1}L\right\} \,\,=\,\,\frac{1}{a_{0}^{2}}\left\{ \boldsymbol{B}\times\frac{\partial\boldsymbol{E}}{\partial t}\right\} _{x}-\frac{1}{a_{0}^{2}}\left\{ \boldsymbol{E}\times\frac{\partial\boldsymbol{B}}{\partial t}\right\} _{x}-\left\{ \boldsymbol{B}\times\left(\boldsymbol{\nabla\times}\boldsymbol{B}\right)\right\} _{x}-\frac{1}{a_{0}^{2}}\left\{ \boldsymbol{E}\times\left(\boldsymbol{\nabla\times}\boldsymbol{E}\right)\right\} _{x}\nonumber \\
+\left\{ \boldsymbol{B}\left(\boldsymbol{\nabla\cdot}\boldsymbol{B}\right)\right\} _{x}+\frac{1}{a_{0}^{2}}\left\{ \boldsymbol{E}\left(\boldsymbol{\nabla\cdot}\boldsymbol{E}\right)\right\} _{x}\,\,\boldsymbol{\longmapsto}\,\,\,\,\left(\text{Real coefficients of \ensuremath{e_{1}}}\right)\,\,,\label{eq:72}\\
\text{Im}\left\{ e_{1}L\right\} \,\,=\,\,-\left\{ \boldsymbol{B}\times\frac{\partial\boldsymbol{B}}{\partial t}\right\} _{x}-\frac{1}{a_{0}^{2}}\left\{ \boldsymbol{E}\times\frac{\partial\boldsymbol{E}}{\partial t}\right\} _{x}-\left\{ \boldsymbol{B}\times\left(\boldsymbol{\nabla\times}\boldsymbol{E}\right)\right\} _{x}+\left\{ \boldsymbol{E}\times\left(\boldsymbol{\nabla\times}\boldsymbol{B}\right)\right\} _{x}\,\nonumber \\
-\left\{ \boldsymbol{E}\left(\boldsymbol{\nabla\cdot}\boldsymbol{B}\right)\right\} _{x}+\left\{ \boldsymbol{B}\left(\boldsymbol{\nabla\cdot}\boldsymbol{E}\right)\right\} _{x}\,\,\boldsymbol{\longmapsto}\,\,\,\,\,\text{\ensuremath{\left(\text{Imaginary coefficients of }\ensuremath{e_{1}}\right)}}\,\,,\label{eq:73}
\end{alignat}
\begin{alignat}{1}
\text{Re}\left\{ e_{2}M\right\} \,\,=\,\,\frac{1}{a_{0}^{2}}\left\{ \boldsymbol{B}\times\frac{\partial\boldsymbol{E}}{\partial t}\right\} _{y}-\frac{1}{a_{0}^{2}}\left\{ \boldsymbol{E}\times\frac{\partial\boldsymbol{B}}{\partial t}\right\} _{y}-\left\{ \boldsymbol{B}\times\left(\boldsymbol{\nabla\times}\boldsymbol{B}\right)\right\} _{y}-\frac{1}{a_{0}^{2}}\left\{ \boldsymbol{E}\times\left(\boldsymbol{\nabla\times}\boldsymbol{E}\right)\right\} _{y}\nonumber \\
+\left\{ \boldsymbol{B}\left(\boldsymbol{\nabla\cdot}\boldsymbol{B}\right)\right\} _{y}+\frac{1}{a_{0}^{2}}\left\{ \boldsymbol{E}\left(\boldsymbol{\nabla\cdot}\boldsymbol{E}\right)\right\} _{y}\,\boldsymbol{\longmapsto}\,\,\,\text{\ensuremath{\left(\text{Real\,coefficients\,of}\ensuremath{\,e_{2}}\right)}}\,\,,\label{eq:74}\\
\text{Im}\left\{ e_{2}M\right\} \,\,=\,\,-\left\{ \boldsymbol{B}\times\frac{\partial\boldsymbol{B}}{\partial t}\right\} _{y}-\frac{1}{a_{0}^{2}}\left\{ \boldsymbol{E}\times\frac{\partial\boldsymbol{E}}{\partial t}\right\} _{y}-\left\{ \boldsymbol{B}\times\left(\boldsymbol{\nabla\times}\boldsymbol{E}\right)\right\} _{y}+\left\{ \boldsymbol{E}\times\left(\boldsymbol{\nabla\times}\boldsymbol{E}\right)\right\} _{y}\,\nonumber \\
-\left\{ \boldsymbol{E}\left(\boldsymbol{\nabla\cdot}\boldsymbol{B}\right)\right\} _{y}+\left\{ \boldsymbol{B}\left(\boldsymbol{\nabla\cdot}\boldsymbol{E}\right)\right\} _{y}\,\,\boldsymbol{\longmapsto}\,\,\text{\ensuremath{\left(\text{Imaginary\,coefficients\,of}\,\ensuremath{e_{2}}\right)}}\,\,,\label{eq:75}
\end{alignat}

\begin{alignat}{1}
\text{Re}\left\{ e_{3}N\right\} \,\,=\,\,\frac{1}{a_{0}^{2}}\left\{ \boldsymbol{B}\times\frac{\partial\boldsymbol{E}}{\partial t}\right\} _{z}-\frac{1}{a_{0}^{2}}\left\{ \boldsymbol{E}\times\frac{\partial\boldsymbol{B}}{\partial t}\right\} _{z}-\left\{ \boldsymbol{B}\times\left(\boldsymbol{\nabla\times}\boldsymbol{B}\right)\right\} _{z}-\frac{1}{a_{0}^{2}}\left\{ \boldsymbol{E}\times\left(\boldsymbol{\nabla\times}\boldsymbol{E}\right)\right\} _{z}\nonumber \\
+\left\{ \boldsymbol{B}\left(\boldsymbol{\nabla\cdot}\boldsymbol{B}\right)\right\} _{z}+\frac{1}{a_{0}^{2}}\left\{ \boldsymbol{E}\left(\boldsymbol{\nabla\cdot}\boldsymbol{E}\right)\right\} _{z}\,\,\boldsymbol{\longmapsto}\,\text{\ensuremath{\left(\text{Real\,coefficients\,of\,}\ensuremath{e_{3}}\right)}}\,\,,\label{eq:76}\\
\text{Im}\left\{ e_{3}N\right\} \,\,=\,\,-\left\{ \boldsymbol{B}\times\frac{\partial\boldsymbol{B}}{\partial t}\right\} _{y}-\frac{1}{a_{0}^{2}}\left\{ \boldsymbol{E}\times\frac{\partial\boldsymbol{E}}{\partial t}\right\} _{y}-\left\{ \boldsymbol{B}\times\left(\boldsymbol{\nabla\times}\boldsymbol{E}\right)\right\} _{y}+\left\{ \boldsymbol{E}\times\left(\boldsymbol{\nabla\times}\boldsymbol{E}\right)\right\} _{y}\,\nonumber \\
-\left\{ \boldsymbol{E}\left(\boldsymbol{\nabla\cdot}\boldsymbol{B}\right)\right\} _{y}+\left\{ \boldsymbol{B}\left(\boldsymbol{\nabla\cdot}\boldsymbol{E}\right)\right\} _{y}\,\boldsymbol{\longmapsto}\left(\text{Imaginary\,cofficients of \,}e_{3}\right)\,\,,\label{eq:77}
\end{alignat}
along with,
\begin{alignat}{1}
\text{Re}\left\{ e_{0}O\right\} \,\,=\,\,-\frac{1}{a_{0}^{2}}\left(\boldsymbol{B}\cdot\frac{\partial\boldsymbol{E}}{\partial t}\right)+\frac{1}{a_{0}^{2}}\left(\boldsymbol{E}\cdot\frac{\partial\boldsymbol{B}}{\partial t}\right)+\boldsymbol{B}\cdot\left(\boldsymbol{\nabla\times B}\right)\nonumber \\
+\frac{1}{a_{0}^{2}}\left\{ \boldsymbol{E}\cdot\left(\boldsymbol{\nabla\times E}\right)\right\} \,\boldsymbol{\longmapsto}\,\, & \left(\text{Real coefficients of \ensuremath{\,e_{0}}}\right)\,\,,\label{eq:78}\\
\text{Im}\left\{ e_{0}O\right\} \,\,=\,\,\left(\boldsymbol{B}\cdot\frac{\partial\boldsymbol{B}}{\partial t}\right)+\frac{1}{a_{0}^{2}}\left(\boldsymbol{E}\cdot\frac{\partial\boldsymbol{E}}{\partial t}\right)+\boldsymbol{B}\cdot\left(\boldsymbol{\nabla\times E}\right)\nonumber \\
-\boldsymbol{E}\cdot\left(\boldsymbol{\nabla\times B}\right)\,\boldsymbol{\longmapsto}\,\, & \left(\text{Imaginary coefficients of\ensuremath{\,e_{0}}}\right)\,\,.\label{eq:79}
\end{alignat}
Similarly, the right hand part of equation (\ref{eq:70}) can also
be expressed in terms of the following quaternionic form
\begin{align}
-\boldsymbol{\bar{\Psi}}\circ\mathbb{J}\,\,\left\{ e_{1},\,e_{2},\,e_{3};\,\,\,e_{0}\right\} \,\,= & \,\,\ensuremath{e_{1}}L'+\ensuremath{e_{2}}M'+\ensuremath{e_{3}}N'+\ensuremath{e_{0}}O'\,\,,\,\,\,\forall\,(L',\,M',\,N',\,O')\in\mathbb{C}\label{eq:80}
\end{align}
where the real and imaginary quaternionic components $(L',M',N',\,\text{and\,}O')$
are
\begin{align}
\text{Re}\left\{ e_{1}L'\right\} \,\,= & \,\,-\mu\left(\boldsymbol{B\times J^{e}}\right)_{x}+\frac{1}{a_{0}^{2}\epsilon}\left(\boldsymbol{E\times J^{\mathfrak{m}}}\right)_{x}+\mu\left(\rho^{\mathfrak{m}}B_{x}\right)\nonumber \\
 & +\frac{1}{a_{0}^{2}\epsilon}\left(\rho^{e}E_{x}\right)\,\,\longmapsto\left(\text{Real cofficients of}\,e_{1}\right)\,\,,\label{eq:81}\\
\text{Im}\left\{ e_{1}L'\right\} \,\,= & \,\,\frac{1}{a_{0}}\biggl\{\mu\left(\boldsymbol{E\times J^{e}}\right)_{x}+\frac{1}{\epsilon}\left(\boldsymbol{B\times J^{m}}\right)_{x}-\mu\left(\rho^{\mathfrak{m}}E_{x}\right)\nonumber \\
 & +\frac{1}{\epsilon}\left(\rho^{e}B_{x}\right)\biggr\}\,\,\longmapsto\left(\text{Imaginary}\text{ coefficients of }\,e_{1}\right)\,\,,\label{eq:82}\\
\text{Re}\left\{ e_{2}M'\right\} \,\,= & \,\,-\mu\left(\boldsymbol{B\times J^{e}}\right)_{y}+\frac{1}{a_{0}^{2}\epsilon}\left(\boldsymbol{E\times J^{\mathfrak{m}}}\right)_{y}+\mu\left(\rho^{\mathfrak{m}}B_{y}\right)\nonumber \\
 & +\frac{1}{a_{0}^{2}\epsilon}\left(\rho^{e}E_{y}\right)\,\,\longmapsto\left(\text{Real cofficients of}\,e_{2}\right)\,\,,\label{eq:83}\\
\text{Im}\left\{ e_{2}M'\right\} \,\,= & \,\,\frac{1}{a_{0}}\biggl\{\mu\left(\boldsymbol{E\times J^{e}}\right)_{y}+\frac{1}{\epsilon}\left(\boldsymbol{B\times J^{m}}\right)_{y}-\mu\left(\rho^{\mathfrak{m}}E_{y}\right)\nonumber \\
 & +\frac{1}{\epsilon}\left(\rho^{e}B_{y}\right)\biggr\}\,\,\longmapsto\left(\text{Imaginary}\text{ coefficients of }\,e_{2}\right)\,\,,\label{eq:84}\\
\text{Re}\left\{ e_{3}N'\right\} \,\,= & \,\,-\mu\left(\boldsymbol{B\times J^{e}}\right)_{z}+\frac{1}{a_{0}^{2}\epsilon}\left(\boldsymbol{E\times J^{\mathfrak{m}}}\right)_{z}+\mu\left(\rho^{\mathfrak{m}}B_{z}\right)\nonumber \\
 & +\frac{1}{a_{0}^{2}\epsilon}\left(\rho^{e}E_{z}\right)\,\,\longmapsto\left(\text{Real cofficients of}\,e_{3}\right)\,\,,\label{eq:85}\\
\text{Im}\left\{ e_{3}N'\right\} \,\,= & \,\,\frac{1}{a_{0}}\biggl\{\mu\left(\boldsymbol{E\times J^{e}}\right)_{z}+\frac{1}{\epsilon}\left(\boldsymbol{B\times J^{m}}\right)_{z}-\mu\left(\rho^{\mathfrak{m}}E_{z}\right)\nonumber \\
 & +\frac{1}{\epsilon}\left(\rho^{e}B_{z}\right)\biggr\}\,\,\longmapsto\left(\text{Imaginary}\text{ coefficients of }\,e_{3}\right)\,\,,\label{eq:86}
\end{align}
and
\begin{align}
\text{Re}\left\{ e_{0}O'\right\} \,\,= & \,\,\mu\left(\boldsymbol{B}\cdot\boldsymbol{J}^{e}\right)-\frac{1}{\epsilon}\left(\boldsymbol{E}\cdot\boldsymbol{J}^{\mathfrak{m}}\right)\,\,\longmapsto\,\,\,\,\,\,\,\left(\text{Real}\text{ coefficients of }\,e_{0}\right)\,\,,\label{eq:87}\\
\text{Im}\left\{ e_{0}O'\right\} \,\,= & \,\,\frac{1}{a_{0}}\left\{ -\frac{1}{\epsilon}\left(\boldsymbol{B}\cdot\boldsymbol{J}^{\mathfrak{m}}\right)-\mu\left(\boldsymbol{E}\cdot\boldsymbol{J}^{e}\right)\right\} \,\,\longmapsto\,\,\,\left(\text{Imaginary}\text{ coefficients of }\,e_{0}\,\right)\,\,.\label{eq:88}
\end{align}
The above quaternionic analysis shows that the left and right-hand
sides of equations (\ref{eq:70}) resemble to one another, if the
quaternionic coefficients $(L,M,N,O)$ and $(L',M',N',O')$ coincide
to each other, i.e.,
\begin{align}
e_{1}L(\text{Re,\,Im})\,\,\cong & \,\,\,e_{1}L'(\text{Re,\,Im})\nonumber \\
e_{2}M(\text{Re,\,Im})\,\,\cong & \,\,\,e_{2}M'(\text{Re,\,Im})\nonumber \\
e_{3}N(\text{Re,\,Im})\,\,\cong & \,\,\,e_{3}N'(\text{Re,\,Im})\nonumber \\
e_{0}O(\text{Re,\,Im})\,\,\cong & \,\,\,e_{0}O'(\text{Re,\,Im})\,\,.\label{eq:89}
\end{align}
At first we would like to equate the scalar components, i.e., $e_{0}O(\text{Re,\,Im})\,\,\cong\,\,\,e_{0}O'(\text{Re,\,Im})$.
For this resemble, we equate the imaginary part of quaternionic scalar
coefficient ($e_{0}$) that gives the conservation of energy to require
the flow of hydro-electromagnetic dyonic plasma as
\begin{alignat}{1}
\boldsymbol{B}\cdot\frac{\partial\boldsymbol{B}}{\partial t}+\frac{1}{a_{0}^{2}}\left(\boldsymbol{E}\cdot\frac{\partial\boldsymbol{E}}{\partial t}\right)+\boldsymbol{B}\cdot\left(\boldsymbol{\nabla\times E}\right)-\boldsymbol{E}\cdot\left(\boldsymbol{\nabla\times B}\right)+\frac{1}{\epsilon}\left(\boldsymbol{B}\cdot\boldsymbol{J}^{\mathfrak{m}}\right)+\mu\left(\boldsymbol{E}\cdot\boldsymbol{J}^{e}\right)\,=\,\, & 0\,,\label{eq:90}
\end{alignat}
which can further reduces as 
\begin{alignat}{1}
\frac{1}{2}\left(\frac{\partial B^{2}}{\partial t}+\frac{1}{a_{0}^{2}}\frac{\partial E^{2}}{\partial t}\right)+\boldsymbol{\nabla}\cdot\left(\boldsymbol{E}\times\boldsymbol{B}\right)+\frac{1}{\epsilon}\left(\boldsymbol{B}\cdot\boldsymbol{J}^{\mathfrak{m}}\right)+\mu\left(\boldsymbol{E}\cdot\boldsymbol{J}^{e}\right)\,=\,\, & 0\,\,.\label{eq:91}
\end{alignat}
Equation (\ref{eq:91}) indicates the energy theorem also known as
the \textit{Poynting's theorem} for the generalized electromagnetic
fluid of dyons, where the first term represents the hydro-electric
and hydro-magnetic fields energy, second term represents the average
energy flux and the third and fourth terms represent the work done
by the field on the magnetic monopoles and electrons. Interestingly,
equation (\ref{eq:91}) shows a resemblance to the \textit{Bernoulli's
theorem} in which we study the conservation of energy to the case
of dyonic-fluid flow. If we equating the real part of quaternionic
unit $e_{0}$ in equation (\ref{eq:70}) the complexfied Dirac-Maxwell
equations for plasma-fluid are obtained.

Now, to find the force equation or conservation of momentum for the
hydro-electromagnetic field of dyonic cold plasma, we proceed by equating
the real coefficients of $e_{j}X_{j}(\text{Re,\,Im})\,\,\cong\,\,\,e_{j}X_{j}'(\text{Re,\,Im})\,,$
$\forall\,j=1,2,3\,\text{and}\,X_{j}\simeq\left(L,M,N\right),$ $X_{j}'\simeq\left(L',M',N'\right)$
as,
\begin{alignat}{1}
\frac{1}{a_{0}^{2}}\left\{ \boldsymbol{B}\times\frac{\partial\boldsymbol{E}}{\partial t}\right\} -\frac{1}{a_{0}^{2}}\left\{ \boldsymbol{E}\times\frac{\partial\boldsymbol{B}}{\partial t}\right\}  & -\left\{ \boldsymbol{B}\times\left(\boldsymbol{\nabla\times}\boldsymbol{B}\right)\right\} -\frac{1}{a_{0}^{2}}\left\{ \boldsymbol{E}\times\left(\boldsymbol{\nabla\times}\boldsymbol{E}\right)\right\} +\left\{ \boldsymbol{B}\left(\boldsymbol{\nabla\cdot}\boldsymbol{B}\right)\right\} \nonumber \\
+\frac{1}{a_{0}^{2}}\left\{ \boldsymbol{E}\left(\boldsymbol{\nabla\cdot}\boldsymbol{E}\right)\right\} \,=\,- & \mu\left(\boldsymbol{B}\times\boldsymbol{J}^{e}\right)+\frac{1}{a_{0}^{2}\epsilon}\left(\boldsymbol{E}\times\boldsymbol{J}^{\mathfrak{m}}\right)+\mu\left(\rho^{\mathfrak{m}}\boldsymbol{B}\right)+\frac{1}{a_{0}^{2}\epsilon}\left(\rho^{e}\boldsymbol{E}\right)\,,\label{eq:92}
\end{alignat}
which simplifies to
\begin{alignat}{1}
-\frac{1}{a_{0}^{2}}\frac{\partial\boldsymbol{\mathcal{H}}}{\partial t}-\frac{1}{2}\boldsymbol{\nabla}\left(B^{2}+\frac{1}{a_{0}^{2}}E^{2}\right)+\left(\boldsymbol{B}\boldsymbol{\cdot\nabla}\right)\boldsymbol{B}+\frac{1}{a_{0}^{2}}\left(\boldsymbol{E}\boldsymbol{\cdot\nabla}\right)\boldsymbol{E}+\boldsymbol{B}\left(\boldsymbol{\nabla\cdot B}\right)+\frac{1}{a_{0}^{2}}\boldsymbol{E}\left(\boldsymbol{\nabla\cdot E}\right)\nonumber \\
=\,\,\,\mu\left(\rho^{m}\boldsymbol{B}\right)+\frac{1}{a_{0}^{2}\epsilon}\left(\rho^{e}\boldsymbol{E}\right)-\mu\left(\boldsymbol{B}\times\boldsymbol{J}^{e}\right)+\frac{1}{a_{0}^{2}\epsilon}\left(\boldsymbol{E}\times\boldsymbol{J}^{m}\right)\,,\label{eq:93}
\end{alignat}
where $\boldsymbol{\mathcal{H}}=\left(\boldsymbol{E}\times\boldsymbol{B}\right)$
represents the fluidic power flux (or fluidic Poynting vector) that
provides the energy transport of plasma fluid by the hydro-electromagnetic
field per unit volume per unit time. Interestingly, equation (\ref{eq:93})
represented the force per unit volume due to the generalized hydro-electromagnetic
energy of dyonic cold plasma, so that
\begin{alignat}{1}
\boldsymbol{\mathcal{F}}\,\,=\,\, & \frac{1}{a_{0}^{2}}\frac{\partial\boldsymbol{\mathcal{H}}}{\partial t}+\frac{1}{2}\boldsymbol{\nabla}\left(B^{2}+\frac{1}{a_{0}^{2}}E^{2}\right)-\left(\boldsymbol{B}\boldsymbol{\cdot\nabla}\right)\boldsymbol{B}-\frac{1}{a_{0}^{2}}\left(\boldsymbol{E}\boldsymbol{\cdot\nabla}\right)\boldsymbol{E}-\left(\boldsymbol{\nabla\cdot B}\right)\boldsymbol{B}\nonumber \\
 & -\frac{1}{a_{0}^{2}}\left(\boldsymbol{\nabla\cdot E}\right)\boldsymbol{E}+\mu\left(\rho^{m}\boldsymbol{B}\right)+\frac{1}{a_{0}^{2}\epsilon}\left(\rho^{e}\boldsymbol{E}\right)-\mu\left(\boldsymbol{B}\times\boldsymbol{J}^{e}\right)+\frac{1}{a_{0}^{2}\epsilon}\left(\boldsymbol{E}\times\boldsymbol{J}^{m}\right)\,\,,\label{eq:94}
\end{alignat}
where $\mathcal{\boldsymbol{\mathcal{F}}}$ is the generalized quaternionic
fluid force per unit volume superimposed by three terms i.e. the stress
tensor, the fluidic power flux and the dynamics of dyonic particles
per unit volume in cold plasma. Moreover, equation (\ref{eq:94})
leads to the following compact form,
\begin{align}
\boldsymbol{\mathcal{F}}\,\,= & \,\,\left(\boldsymbol{\nabla}\cdot\boldsymbol{\overleftrightarrow{T}}\right)+\boldsymbol{F}_{ff}+\boldsymbol{F}_{dyons}\,,\label{eq:95}
\end{align}
where the divergence of viscous stress tensor which acts analogous
to the Maxwell stress tensor yields
\begin{align}
\boldsymbol{\nabla\cdot}\boldsymbol{\overleftrightarrow{T}}\,\,=- & \frac{1}{a_{0}^{2}}\left[\left(\boldsymbol{\nabla\cdot E}\right)\boldsymbol{E}+\left(\boldsymbol{E}\boldsymbol{\cdot\nabla}\right)\boldsymbol{E}-\frac{1}{2}\boldsymbol{\nabla}E^{2}\right]-\left[\left(\boldsymbol{B}\boldsymbol{\cdot\nabla}\right)\boldsymbol{B}+\left(\boldsymbol{\nabla\cdot B}\right)\boldsymbol{B}-\frac{1}{2}\boldsymbol{\nabla}B^{2}\right]\,\,,\label{eq:96}
\end{align}
and the forces arises due to the fluidic power flux ($\boldsymbol{F}_{ff}$)
and due to electromagnetic dyonic fluid particles ($\boldsymbol{F}_{dyons}$)
become
\begin{align}
\boldsymbol{F}_{ff}\,\, & \simeq\,\,\frac{1}{a_{0}^{2}}\frac{\partial\boldsymbol{\mathcal{H}}}{\partial t}\,\,,\label{eq:97}\\
\boldsymbol{F}_{dyons}\,\, & \simeq\,\,\mu\left(\rho^{m}\boldsymbol{B}\right)-\frac{1}{a_{0}^{2}\epsilon}\rho^{e}\boldsymbol{E}-\mu\left(\boldsymbol{B}\times\boldsymbol{J}^{e}\right)+\frac{1}{a_{0}^{2}\epsilon}\left(\boldsymbol{E}\times\boldsymbol{J}^{m}\right)\,.\label{eq:98}
\end{align}
Thus the obtained equation (\ref{eq:95}) represents the quaternionic
generalization of Navier-Stokes like equation in case of dyonic cold
plasma. We also can write the simplified form of Navier-Stokes like
equation if we put the value of quaternionic fluid force, i.e., 
\begin{align}
\boldsymbol{\mathcal{F}}\,\,\simeq\,\,\rho\left(\frac{\partial}{\partial t}+\boldsymbol{v}\cdot\boldsymbol{\nabla}\right)\boldsymbol{v}\,\, & =\,\,\,\left(\boldsymbol{\nabla}\cdot\boldsymbol{\overleftrightarrow{T}}\right)+\boldsymbol{F}_{ff}+\boldsymbol{F}_{dyons}\,.\label{eq:99}
\end{align}
Therefore, by combing the above Navier-Stokes like equation (\ref{eq:99})
with Dirac-Maxwell equations (\ref{eq:68})-(\ref{eq:69}), the resultant
MHD fluid equations provide a complete description of dyonic cold
plasma. In order to obtain the conservation law for fluid momentum,
we may write the equation (\ref{eq:99}) in terms of linear momentum
as,
\begin{alignat}{1}
\frac{\partial\boldsymbol{P}_{\text{mech}}}{\partial t}\,\,= & \,\left(\boldsymbol{\nabla}\cdot\boldsymbol{\overleftrightarrow{T}}\right)+\frac{\partial\boldsymbol{P}_{\text{hydroem}}}{\partial t}\,,\label{eq:100}
\end{alignat}
where $\boldsymbol{P}_{\text{mech}}$ represents the mechanical momentum
and $\boldsymbol{P}_{\text{hydroem}}$ represents the total generalized
hydro-electromagnetic momentum of dyonic cold plasma. Here we define
the total generalized hydro-electromagnetic force as $\boldsymbol{F}_{\text{hydroem}}\,=\,\left(\boldsymbol{F}_{ff}+\boldsymbol{F}_{dyons}\right)\,\simeq\,\partial\boldsymbol{P}_{\text{hydroem}}/\partial t$.
Therefore, we get
\begin{alignat}{1}
\frac{\partial\mathbb{G}}{\partial t}+\boldsymbol{\nabla}\cdot\boldsymbol{\overleftrightarrow{T}}\,\,=\, & 0\,,\label{eq:101}
\end{alignat}
where the resultant momentum becomes$\mathbb{G}\rightarrow\left(\boldsymbol{P}_{\text{hydroem}}-\boldsymbol{P}_{\text{mech}}\right).$
Equation (\ref{eq:101}) represented generalized continuity equation
for the case of generalized hydro-electromagnetic fluid-momentum,
where the viscous stress tensor $\boldsymbol{\overleftrightarrow{T}}$
works as the source current and the term $\mathbb{G}$ works as the
source density of the system. Correspondingly, if we equate the imaginary
coefficient of quaternionic unit $e_{j}$ $\left(\forall\,j=1,2,3\right)$
in equation (\ref{eq:89}) we obtain again the complexfied Dirac-Maxwell
like equations. Thus the interesting part of our present quaternionic
formalism is the generalized energy- momentum conservation of the
hydro-electromagnetic fluid of dyonic cold plasma shows the invariant
nature under the duality and Lorentz transformations.

\section{Quaternionic wave equations for cold plasma fluid}

In this section, we shall describe the wave equations of electromagnetic
fluid plasma consisting with dyons. To obtain the dyonic wave equations
for cold plasma fluid, we may operate $\mathbb{D}$ by left on the
quaternionic field equation as
\begin{alignat}{1}
\mathbb{D}\circ(\mathbb{\bar{D}}\circ\boldsymbol{\Psi})\,\,= & \,-\mathbb{D}\circ\mathbb{J}\,.\label{eq:102}
\end{alignat}
The quaternionic expression for the left hand side of equation (\ref{eq:102})
becomes
\begin{align}
\mathbb{D}\circ(\mathbb{\bar{D}}\circ\boldsymbol{\Psi})\,\,\left\{ e_{1},\,e_{2},\,e_{3};\,\,\,e_{0}\right\} \,\,= & \,\,\ensuremath{e_{1}}P+\ensuremath{e_{2}}Q+\ensuremath{e_{3}}R+\ensuremath{e_{0}}S\,\,,\,\,\,\forall\,(P,\,Q,\,R,\,S)\in\mathbb{C}\label{eq:103}
\end{align}
where the real and imaginary components of quaternionic coefficient
$(P,\,Q,\,R,\,S)$ are
\begin{alignat}{1}
\text{Re}\left\{ e_{1}P\right\} \,\,= & \,\,\left(\frac{\partial^{2}B_{x}}{\partial x^{2}}-\frac{1}{a_{0}^{2}}\frac{\partial^{2}B_{x}}{\partial t^{2}}\right)\,,\,\,\,\,\text{Im}\left\{ e_{1}P\right\} \,\,=\,\,\frac{1}{a_{0}}\left(\frac{\partial^{2}E_{x}}{\partial x^{2}}-\frac{1}{a_{0}^{2}}\frac{\partial^{2}E_{x}}{\partial t^{2}}\right)\,,\nonumber \\
\text{Re}\left\{ e_{2}Q\right\} \,\,= & \,\,\left(\frac{\partial^{2}B_{y}}{\partial y^{2}}-\frac{1}{a_{0}^{2}}\frac{\partial^{2}B_{y}}{\partial t^{2}}\right)\,,\,\,\,\,\text{Im}\left\{ e_{2}Q\right\} \,\,=\,\,\frac{1}{a_{0}}\left(\frac{\partial^{2}E_{y}}{\partial y^{2}}-\frac{1}{a_{0}^{2}}\frac{\partial^{2}E_{y}}{\partial t^{2}}\right)\,,\nonumber \\
\text{Re}\left\{ e_{3}R\right\} \,\,= & \,\,\left(\frac{\partial^{2}B_{z}}{\partial z^{2}}-\frac{1}{a_{0}^{2}}\frac{\partial^{2}B_{z}}{\partial t^{2}}\right)\,,\,\,\,\,\text{Im}\left\{ e_{2}R\right\} \,\,=\,\,\frac{1}{a_{0}}\left(\frac{\partial^{2}E_{z}}{\partial z^{2}}-\frac{1}{a_{0}^{2}}\frac{\partial^{2}E_{z}}{\partial t^{2}}\right)\,,\nonumber \\
\text{Re}\left\{ e_{0}S\right\} \,\,= & 0\,,\,\,\,\,\text{Im}\left\{ e_{0}S\right\} \,\,=\,\,0\,\,.\label{eq:104}
\end{alignat}
Equation (\ref{eq:104}) associated classical wave equation for hydro-electric
and hydro-magnetic fields without containing any source. Correspondingly,
the quaternionic source expression for the right hand side of equation
(\ref{eq:102}) can be written as
\begin{align}
-\mathbb{D}\circ\mathbb{J}\,\,\left\{ e_{1},\,e_{2},\,e_{3};\,\,\,e_{0}\right\} \,\,= & \,\,\ensuremath{e_{1}}P'+\ensuremath{e_{2}}Q'+\ensuremath{e_{3}}R'+\ensuremath{e_{0}}S'\,,\,\,\,\forall\,(P',\,Q',\,R',\,S')\in\mathbb{C}\label{eq:105}
\end{align}
where the real and imaginary components are
\begin{alignat}{1}
\text{Re}\left\{ e_{1}P'\right\} \,\,= & \,\,\mu\left(\frac{\partial J_{z}^{e}}{\partial y}-\frac{\partial J_{y}^{e}}{\partial z}-\frac{1}{a_{0}^{2}\mu\epsilon}\frac{\partial J_{x}^{\mathfrak{m}}}{\partial t}-\frac{\partial\rho^{\mathfrak{m}}}{\partial x}\right)\,\,,\nonumber \\
\text{Im}\left\{ e_{1}P'\right\} \,\,= & -\frac{1}{a_{0}\epsilon}\left(\frac{\partial J_{z}^{\mathfrak{m}}}{\partial y}-\frac{\partial J_{y}^{\mathfrak{m}}}{\partial z}+\mu\epsilon\frac{\partial J_{x}^{e}}{\partial t}+\frac{\partial\rho^{e}}{\partial x}\right)\,\,,\nonumber \\
\text{Re}\left\{ e_{2}Q'\right\} \,\,= & \,\,\mu\left(\frac{\partial J_{x}^{e}}{\partial z}-\frac{\partial J_{z}^{e}}{\partial x}-\frac{1}{a_{0}^{2}\mu\epsilon}\frac{\partial J_{y}^{\mathfrak{m}}}{\partial t}-\frac{\partial\rho^{\mathfrak{m}}}{\partial y}\right)\,\,,\nonumber \\
\text{Im}\left\{ e_{2}Q'\right\} \,\,= & -\frac{1}{a_{0}\epsilon}\left(\frac{\partial J_{x}^{\mathfrak{m}}}{\partial z}-\frac{\partial J_{z}^{\mathfrak{m}}}{\partial x}+\mu\epsilon\frac{\partial J_{y}^{e}}{\partial t}+\frac{\partial\rho^{e}}{\partial y}\right)\,\,,\nonumber \\
\text{Re}\left\{ e_{3}R'\right\} \,\,= & \,\,\mu\left(\frac{\partial J_{y}^{e}}{\partial x}-\frac{\partial J_{x}^{e}}{\partial y}-\frac{1}{a_{0}^{2}\mu\epsilon}\frac{\partial J_{z}^{\mathfrak{m}}}{\partial t}-\frac{\partial\rho^{\mathfrak{m}}}{\partial z}\right)\,\,,\nonumber \\
\text{Im}\left\{ e_{3}R'\right\} \,\,= & -\frac{1}{a_{0}\epsilon}\left(\frac{\partial J_{y}^{\mathfrak{m}}}{\partial x}-\frac{\partial J_{x}^{\mathfrak{m}}}{\partial y}+\mu\epsilon\frac{\partial J_{z}^{e}}{\partial t}+\frac{\partial\rho^{e}}{\partial z}\right)\,\,,\nonumber \\
\text{Re}\left\{ e_{0}S'\right\} \,\,= & \,\,-\mu\left(\frac{\partial J_{x}^{e}}{\partial x}+\frac{\partial J_{y}^{e}}{\partial y}+\frac{\partial J_{z}^{e}}{\partial z}+\mu\epsilon\frac{\partial\rho^{e}}{\partial t}\right)\,\,,\nonumber \\
\text{Im}\left\{ e_{0}S'\right\} \,\,= & \,\,\frac{1}{a_{0}\epsilon}\left(\frac{\partial J_{x}^{\mathfrak{m}}}{\partial x}+\frac{\partial J_{y}^{\mathfrak{m}}}{\partial y}+\frac{\partial J_{z}^{\mathfrak{m}}}{\partial z}+\frac{\partial\rho^{\mathfrak{m}}}{\partial t}\right)\,\,.\label{eq:106}
\end{alignat}
The physical significant of quaternionic analysis occurs if the left
and right-hand sides of equations (\ref{eq:102}) resemble to one
another, and the quaternionic coefficients $(P,\,Q\,,R\,,S)$ and
$(P'\,,Q'\,,R'\,,S')$ coincide to each other,
\begin{align}
e_{1}P(\text{Re,\,Im})\,\,\cong & \,\,\,e_{1}P'(\text{Re,\,Im})\nonumber \\
e_{2}Q(\text{Re,\,Im})\,\,\cong & \,\,\,e_{2}Q'(\text{Re,\,Im})\nonumber \\
e_{3}R(\text{Re,\,Im})\,\,\cong & \,\,\,e_{3}R'(\text{Re,\,Im})\nonumber \\
e_{0}S(\text{Re,\,Im})\,\,\cong & \,\,\,e_{0}S'(\text{Re,\,Im})\,\,.\label{eq:107}
\end{align}
Now, we may equate the real and imaginary parts of $e_{0}S(\text{Re,\,Im})\,\,\cong\,\,\,e_{0}S'(\text{Re,\,Im})$
given in equation (\ref{eq:107}) and obtained
\begin{alignat}{1}
\boldsymbol{\nabla}\cdot\boldsymbol{J}^{e}+\frac{1}{a_{0}^{2}}\frac{\partial\rho^{e}}{\partial t} & \,\,=\,\,0\,,\label{eq:108}\\
\boldsymbol{\nabla}\cdot\boldsymbol{J}^{\mathfrak{m}}+\frac{\partial\rho^{\mathfrak{m}}}{\partial t} & \,\,=\,\,0\,.\label{eq:109}
\end{alignat}
These equations represented the continuity equations or simply called
conservation of electric and magnetic charges for the dynamics of
cold electrons and cold magnetic-monopoles in dyonic plasma. Therefore,
we obtain the Lorenz gauge like conditions for compressible cold plasma
fluid ($\boldsymbol{\nabla}\cdot\boldsymbol{J}^{D}\neq0$), i.e.
\begin{alignat}{1}
\boldsymbol{\nabla}\cdot\left(q^{e}n^{e}\boldsymbol{u}\right)+\frac{1}{a_{0}^{2}}\frac{\partial(q^{e}n^{e}h)}{\partial t} & \,\,=\,\,0\,,\label{eq:110}\\
\boldsymbol{\nabla}\cdot\left(q^{\mathfrak{m}}n^{\mathfrak{m}}\boldsymbol{v}\right)+\frac{\partial(q^{\mathfrak{m}}n^{\mathfrak{m}}k)}{\partial t} & \,\,=\,\,0\,.\label{eq:111}
\end{alignat}
Correspondingly, from equation (\ref{eq:107}) equating the coefficients
of $e_{j}Y_{j}(\text{Re,\,Im})\,\,\cong\,\,\,e_{j}Y_{j}'(\text{Re,\,Im})\,,$
$\forall\,j=1,2,3\,\text{and}\,Y_{j}\simeq\left(P,Q,R\right),$ $Y_{j}'\simeq\left(P',Q',R'\right)$
as,

\begin{alignat}{1}
\boldsymbol{\nabla}^{2}\boldsymbol{E}-\frac{1}{a_{0}^{2}}\frac{\partial^{2}\boldsymbol{E}}{\partial t^{2}}-\frac{1}{\epsilon}\left(\boldsymbol{\nabla}\rho^{e}\right)-\mu\frac{\partial\boldsymbol{J}^{e}}{\partial t}-\frac{1}{\epsilon}\left(\boldsymbol{\nabla}\times\boldsymbol{J}^{\mathfrak{m}}\right)\,\, & =\,\,0\,,\label{eq:112}\\
\boldsymbol{\nabla}^{2}\boldsymbol{B}-\frac{1}{a_{0}^{2}}\frac{\partial^{2}\boldsymbol{B}}{\partial t^{2}}-\mu\left(\boldsymbol{\nabla}\rho^{\mathfrak{m}}\right)-\frac{1}{a_{0}^{2}\epsilon}\frac{\partial\boldsymbol{J}^{\mathfrak{m}}}{\partial t}+\mu\left(\boldsymbol{\nabla}\times\boldsymbol{J}^{e}\right)\,\, & =\,\,0\,,\label{eq:113}
\end{alignat}
Equations (\ref{eq:112}) and (\ref{eq:113}) represented the generalized
hydro-electric wave and hydro-magnetic wave equations for cold electrons
and cold magnetic-monopoles traveling in dyonic plasma-fluid. On the
other hand, the hydro-electromagnetic wave components can also be
expressed as
\begin{align}
\boldsymbol{\nabla}^{2}\boldsymbol{E}-\frac{1}{a_{0}^{2}}\frac{\partial^{2}\boldsymbol{E}}{\partial t^{2}}-\frac{1}{\epsilon}\left(\boldsymbol{\nabla}(q^{e}n^{e}h)\right)-\mu\frac{\partial\left(q^{e}n^{e}\boldsymbol{u}\right)}{\partial t}-\frac{1}{\epsilon}\left(\boldsymbol{\nabla}\times\left(q^{\mathfrak{m}}n^{\mathfrak{m}}\boldsymbol{v}\right)\right)\,\, & =\,\,0\,,\label{eq:114}\\
\boldsymbol{\nabla}^{2}\boldsymbol{B}-\frac{1}{a_{0}^{2}}\frac{\partial^{2}\boldsymbol{B}}{\partial t^{2}}-\mu\left(\boldsymbol{\nabla}(q^{\mathfrak{m}}n^{\mathfrak{m}}k)\right)-\frac{1}{a_{0}^{2}\epsilon}\frac{\partial\left(q^{\mathfrak{m}}n^{\mathfrak{m}}\boldsymbol{v}\right)}{\partial t}+\mu\left(\boldsymbol{\nabla}\times\left(q^{e}n^{e}\boldsymbol{u}\right)\right)\,\, & =\,\,0\,.\label{eq:115}
\end{align}
In vacuum, equations (\ref{eq:114}) and (\ref{eq:115}) behave like
as free hydro-electromagnetic wave components of cold plasma, i.e.,
\begin{align}
\square\boldsymbol{E} & \,\,=\,\,0\,,\,\,\,\,\,\,\,\text{and}\,\,\,\,\,\square\boldsymbol{B}\,\,=\,\,0\,.\label{eq:116}
\end{align}
However, we may consider dyonic fluid as the two-fluid theory in which
both electrons and magnetic-monopoles propagate through cold plasma-fluid.
Here, two types of wave propagation seem to be theoretically possible,
first wave propagation of electrons and second wave propagation of
magnetic monopoles where we may consider that electrons wave propagation
are too rapid from the magnetic monopoles due to their mass densities.
In following cases we shall discuss the electrons plasma waves and
magnetic-monopoles plasma waves for dyonic fluid propagation.

\paragraph{Case-1 Langmuir like wave propagation:}

Suppose the magnetic monopoles are infinitely massive, so that they
do not contribute to the given fluid motion \cite{key-38}. In this
situation, the whole process of the plasma fluid depends on the electron
inertia. Thus, in this electrons wave or Langmuir like wave propagation
we assume that the initial condition is a unmagnetized cold plasma
fluid containing no source of magnetic monopoles. Then the equation
of motion for electrons cold plasma fluid becomes
\begin{align}
m^{e}n^{e}\left(\frac{\partial}{\partial t}+\boldsymbol{u}\cdot\boldsymbol{\nabla}\right)\boldsymbol{u}\,\,= & \,\,\boldsymbol{F}^{e}\,,\label{eq:117}
\end{align}
where $\boldsymbol{F}^{e}$ represented Lorentz electric force due
to electrons. The electron continuity equation yields 
\begin{align}
\boldsymbol{\nabla}\cdot\left(n^{e}\boldsymbol{u}\right)+\frac{1}{a_{0}^{2}}\frac{\partial(n^{e}h)}{\partial t} & \,\,=\,\,0\,.\label{eq:118}
\end{align}
Correspondingly, the electrons cold plasma fluid also satisfy the
following Maxwell equations
\begin{align}
\boldsymbol{\nabla\cdot E}\, & =\,\frac{\rho^{e}\,}{\epsilon},\nonumber \\
\boldsymbol{\nabla\cdot B}\, & =\,0\,,\nonumber \\
\boldsymbol{\nabla\times E} & \,=-\frac{\partial\boldsymbol{B}}{\partial t}\,,\nonumber \\
\boldsymbol{\nabla\times B}\, & =\,\frac{1}{a_{0}^{2}}\frac{\partial\boldsymbol{E}}{\partial t}+\mu\boldsymbol{J}^{e}\,.\label{eq:119}
\end{align}
As such the hydro-electromagnetic wave equations for electrons-fluid
plasma can be expressed as

\begin{alignat}{1}
\boldsymbol{\nabla}^{2}\boldsymbol{E}-\frac{1}{a_{0}^{2}}\frac{\partial^{2}\boldsymbol{E}}{\partial t^{2}}-\frac{1}{\epsilon}\left(\boldsymbol{\nabla}\rho^{e}\right)-\mu\frac{\partial\boldsymbol{J}^{e}}{\partial t}\,\, & =\,\,0\,,\label{eq:120}\\
\boldsymbol{\nabla}^{2}\boldsymbol{B}-\frac{1}{a_{0}^{2}}\frac{\partial^{2}\boldsymbol{B}}{\partial t^{2}}+\mu\left(\boldsymbol{\nabla}\times\boldsymbol{J}^{e}\right)\,\, & =\,\,0\,.\label{eq:121}
\end{alignat}
These equations are not invariant under duality transformation because
we considering only electrons-fluid plasma. In this case, the generalized
hydro-electromagnetic wave propagation for electrons-fluid plasma
becomes
\begin{align}
\boldsymbol{\nabla}^{2}\Psi-\frac{1}{a_{0}^{2}}\frac{\partial^{2}\Psi}{\partial t^{2}}-\frac{1}{\epsilon}\left(\boldsymbol{\nabla}\rho^{e}\right)-\mu\left[\frac{\partial\boldsymbol{J}^{e}}{\partial t}-\left(\boldsymbol{\nabla}\times\boldsymbol{J}^{e}\right)\right]\,\, & =\,\,0\,.\label{eq:122}
\end{align}

\paragraph{Case-2 \textquoteright t Hooft Polyakov Monopole like wave propagation:}

In \textquoteright t Hooft-Polyakov model \cite{key-39,key-40}, after
symmetry breaking we may find the $U(1)$ gauge theory which shows
the characteristics of Maxwell\textquoteright s electromagnetic theory.
Generally, the \textquoteright t Hooft-Polyakov magnetic monopole
carries one Dirac unit of magnetic charge. Suppose $m^{\mathfrak{m}}$
represented the mass of \textquoteright t Hooft-Polyakov magnetic
monopoles, and for the pure magnetic monopoles-fluid plasma we neglect
the electrons motion ($\rho^{e},\,\boldsymbol{J}^{e}\simeq0$). Then,
the equation of motion for compressible magnetic monopoles-fluid plasma
becomes
\begin{align}
m^{\text{\ensuremath{\mathfrak{m}}}}n^{\mathfrak{m}}\left(\frac{\partial}{\partial t}+\boldsymbol{v}\cdot\boldsymbol{\nabla}\right)\boldsymbol{v}\,\,\,= & \,\,\boldsymbol{F}^{\text{\ensuremath{\mathfrak{m}}}}\,,\label{eq:123}
\end{align}
along with the continuity equation
\begin{align}
\boldsymbol{\nabla}\cdot\left(n^{\text{\ensuremath{\mathfrak{m}}}}\boldsymbol{v}\right)+\frac{1}{a_{0}^{2}}\frac{\partial(n^{\text{\ensuremath{\mathfrak{m}}}}k)}{\partial t} & \,\,=\,\,0\,,\label{eq:124}
\end{align}
where $\boldsymbol{F}^{\text{\ensuremath{\mathfrak{m}}}}$ is Lorentz
magnetic force. The magnetic monopole-fluid satisfy the following
Maxwell equations
\begin{align}
\boldsymbol{\nabla\cdot E}\, & =\,0\,,\nonumber \\
\boldsymbol{\nabla\cdot B}\, & =\,\mu\rho^{\text{\ensuremath{\mathfrak{m}}}}\,,\nonumber \\
\boldsymbol{\nabla\times E} & \,=-\frac{\partial\boldsymbol{B}}{\partial t}-\frac{1}{\epsilon}\boldsymbol{J}^{\mathfrak{m}}\nonumber \\
\boldsymbol{\nabla\times B}\, & =\,\frac{\partial\boldsymbol{E}}{\partial t}\,.\label{eq:125}
\end{align}
Therefore the hydro-electromagnetic wave equations for magnetic monopole
fluid plasma propagation can be written as

\begin{alignat}{1}
\boldsymbol{\nabla}^{2}\boldsymbol{E}-\frac{1}{a_{0}^{2}}\frac{\partial^{2}\boldsymbol{E}}{\partial t^{2}}-\frac{1}{\epsilon}\left(\boldsymbol{\nabla}\times\boldsymbol{J}^{\mathfrak{m}}\right)\,\, & =\,\,0\,,\label{eq:126}\\
\boldsymbol{\nabla}^{2}\boldsymbol{B}-\frac{1}{a_{0}^{2}}\frac{\partial^{2}\boldsymbol{B}}{\partial t^{2}}-\mu\left(\boldsymbol{\nabla}\rho^{\mathfrak{m}}\right)-\frac{1}{a_{0}^{2}\epsilon}\frac{\partial\boldsymbol{J}^{\mathfrak{m}}}{\partial t}\,\, & =\,\,0\,.\label{eq:127}
\end{alignat}
The generalized hydro-electromagnetic wave propagation of magnetic
monopole fluid plasma becomes
\begin{align}
\boldsymbol{\nabla}^{2}\Psi-\frac{1}{a_{0}^{2}}\frac{\partial^{2}\Psi}{\partial t^{2}}-\mu\left(\boldsymbol{\nabla}\rho^{\mathfrak{m}}\right)-\frac{1}{\epsilon}\left[\frac{1}{a_{0}^{2}}\frac{\partial\boldsymbol{J}^{\mathfrak{m}}}{\partial t}+\left(\boldsymbol{\nabla}\times\boldsymbol{J}^{\mathfrak{m}}\right)\right]\,\, & =\,\,0\,.\label{eq:128}
\end{align}
Moreover, in $^{\shortmid}$t Hooft -Polyakov field, the dynamics
of magnetic monopoles having a definite size inside of which massive
fields play a role in providing a smooth structure and outside they
rapidly vanish leaving the field configuration identical to Dirac's
monopoles. A stable monopole solution satisfying Bogomonly condition
in$^{\shortmid}$t Hooft-Polyakov field introduced by Bogomonly-Prasad-Sommerfield
(BPS) \cite{key-41}.

\section{Conclusion}

\noindent The Navier-Stokes equation generally describes a balance
equation to the motion of compressible fluid together with Newton's
second law. There is an important role of Navier-Stokes equation in
MHD, i.e., the MHD equations are the combination of the Navier-Stokes
equation of fluid dynamics and Maxwell's equations of electrodynamics.
In this paper, we have discussed both the Navier-Stokes and Maxwell's
equations for a complete formulation of dual-MHD equations of dyonic
cold plasma-fluid. The dyons existed in cold plasma (where we assume
negligible plasma temperature), are high energetic soliton particles
consisting electrons as well as magnetic monopoles. We have used the
four-dimensional Hamilton algebra to analysis the dynamics of dyonic
cold plasma fluid. The benefit of the quaternionic algebra is that,
it explains both scalar and vector fields in a single frame called
four-vector formulation in Euclidean space-time. Thus, we have described
quaternionic four-velocities, generalized Lamb \& vorticity fields
components, four-current sources, etc. for dyonic cold plasma. We
have expressed the generalized quaternionic hydro-electromagnetic
field that unify the analogy of Lamb-vorticity fields for dyonic cold
plasma fluid. The scalar component of quaternionic hydro-electromagnetic
field has identified to the dual Lorentz-gauge like conditions. We
have derived the generalized quaternionic Dirac-Maxwell like equations
for the conducting electromagnetic fluid of dyonic cold plasma. In
section-5, the generalized Navier-Stokes equation for dyonic cold
plasma fluid has been discussed. We have obtained the generalized
quaternionic form of conservation of energy for hydro-electromagnetic
field by equating the imaginary part of quaternionic scalar coefficient.
The quaternionic form of energy conservation equation is correlated
with the Bernoulli's theorem for dynamics of dyonic plasma fluid.
The real part of quaternionic coefficient represents the generalized
quaternionic Navier-Stokes like equation for dyonic cold plasma fluid.
It is defined that the total amount of forces per unit volume acted
on hydro-electric and hydro-magnetic fields of dyonic cold plasma.
On the other hand, the generalized quaternionic Navier-Stokes equation
may also be identical to the conservation of linear momentum in the
field of dyonic cold plasma. The conservation of linear momentum for
conducting plasma fluid represented the generalized continuity equation
given by equation (\ref{eq:101}). Therefore, the combination of generalized
Dirac-Maxwell equations and the Navier-Stokes equation provided a
complete description of quaternionic dual-MHD equations. In section-6,
we discussed the wave propagation of dyons in generalized hydro-electromagnetic
fields of cold plasma. Conservation of electric, and magnetic charges
with the dynamics of electrons and magnetic-monopoles in conducting
cold plasma fluid has been analyzed. Equations (\ref{eq:112}) and
(\ref{eq:113}) are described the generalized hydro-electric and hydro-magnetic
wave equations for respectively cold electrons and cold magnetic-monopoles
moving in dyonic plasma-fluid. Interestingly, the quaternionic formalism
for dyonic plasma waves emphasized that theoretically there are two
types of waves propagation namely the wave propagation due to electrons,
and the wave propagation due to magnetic monopoles. The electrons
wave propagation are too rapid from the magnetic monopoles due to
their mass densities. Therefore, our present theory predicted that
there have existed the electrons wave (Langmuir like waves) and the
magnetic monopoles wave (\textquoteright t Hooft-Polyakov waves) for
dynamics of dyonic compressible plasma fluid. The generalized Langmuir-\textquoteright t
Hooft-Polyakov wave propagation for electrons and magnetic monopoles-fluid
have been given by equation (\ref{eq:122}) and (\ref{eq:128}).

\noindent On the other side, in experimental point of view there may
be three categories to search the magnetic monopoles (or dyons), viz.
(a) from accelerator searches (b) from direct searches and (c) from
astrophysical bounds. For accelerator searchers, the magnetic monopoles
should be produced in particle accelerator experiments if the collision
energy is sufficiently high, i.e., higher than $2Mc^{2}$. In order
to check for GUT monopoles, the required energy is at least 12 orders
of magnitude higher than the energies available at the Large Hadron
Collider (LHC). Therefore, it is unrealistic to expect that they could
be produced in any foreseeable particle accelerators. Except to produce
magnetic monopoles in an experiment, one can also try to look for
monopoles that already exist in the universe. Since, the monopoles
are stable particles, therefore monopoles created in the early universe
should still be around. Because of the Dirac quantization condition,
their magnetic field is strong, and their behavior is very different
from other, electrically charged particles. In astrophysical bounds,
the magnetic monopoles would also have astrophysical effects, which
can be used to look for them and constrain their flux. Therefore,
experimentally it is very tough to detect magnetic monopoles (or dyons)
due to its huge energy.
\begin{description}
\item [{\textbf{Acknowledgments:}}] The authors would like to thank the
anonymous reviewers for their helpful and constructive comments that
greatly contributed to improving the final version of the paper.
\end{description}

\end{document}